\documentclass{article}
\usepackage{setspace}

\usepackage{PRIMEarxiv}

\usepackage[utf8]{inputenc} 
\usepackage[T1]{fontenc}    
\usepackage{url}            
\usepackage{booktabs}       
\usepackage{amsfonts}       
\usepackage{nicefrac}       
\usepackage{microtype}      
\usepackage{lipsum}
\usepackage{fancyhdr}       
\usepackage{graphicx}       
\usepackage{bm}
\usepackage{multirow}
\usepackage{adjustbox}
\usepackage{float}
\usepackage{parskip}
\usepackage{algorithm,algpseudocode}
\usepackage{comment}
\usepackage{subcaption}  

\graphicspath{{media/}}     
\RequirePackage{amsthm,amsmath,amsfonts,amssymb}
\RequirePackage[round,authoryear]{natbib}
\RequirePackage[colorlinks,citecolor=blue,urlcolor=blue,linkcolor=blue]{hyperref}

\title{\bf A Statistician's Overview of Physics-Informed Neural Networks for Spatio-Temporal Data
}

\author{
  Christopher K. Wikle\thanks{Corresponding author} \\
  University of Missouri \\
  \texttt{wiklec@missouri.edu} \\
   \And
  Joshua North \\
  Lawrence Berkeley National Laboratory \\
  \texttt{jsnorth@lbl.gov} \\
   \And
  Giri Gopalan \\
  Los Alamos National Laboratory\\
  \texttt{ggopalan@lanl.gov} \\
   \And
  Myungsoo Yoo  \\
  The University of Texas at Austin\\
  \texttt{myungsoo.yoo@austin.utexas.edu} \\
}

\begin{document}
\maketitle

\begin{abstract}
The recent success of deep neural network models with physical constraints (so-called, Physics-Informed Neural Networks, PINNs) has led to renewed interest in the incorporation of mechanistic information in predictive models.  Statisticians and others have long been interested in this problem, which has led to  several practical and innovative solutions dating back decades.  In this overview, we focus on the problem of data-driven prediction and inference of dynamic spatio-temporal processes that include mechanistic information, such as would be available from partial differential equations, with a strong focus on the quantification of uncertainty associated with data, process, and parameters. We give a brief review of several paradigms and focus our attention on Bayesian implementations given they naturally accommodate uncertainty quantification. We then show that it is straight-forward to include the Bayesian PINN (B-PINN) within the Bayesian hierarchical model (BHM) framework that has long been considered for modeling dynamic spatio-temporal processes. Such a BHM-PINN is illustrated via a simulation study in which a latent nonlinear Burgers' equation PDE governs the dynamics of Poisson distributed spatio-temporal data. 
\end{abstract}


\section{Introduction}
\label{sec:intro}

Paraphrasing Galileo, the laws of nature are written in the language of mathematics. Indeed, much of our understanding of the world around us, including physical laws and principles, is codified in terms of mechanistic formulas, such as partial and ordinary differential equations. From a statistical perspective, one can attempt to infer mechanistic equations that describe the system under study. Moreover, one can use {\it a priori} knowledge of mechanistic equations describing a physical process of interest, along with data, for inference (e.g., learning the parameters that control the specified equations or initial conditions) or prediction (e.g., forecasting/interpolating).  

When data are involved, one inherently has uncertainty when building models and optimization methods become an integral part of understanding the world around us. The classic example is when Carl Friedrich Gauss used the method of least squares in the early 1800s to successfully calculate the orbits of celestial bodies; importantly, Gauss used Kepler's laws of celestial mechanics as constraints to his least squares solution \citep{gauss1809theoria}.  Indeed, statisticians have long been interested in using proposed mathematical relationships in the presence of observations to motivate statistical models. For example, as described in \citet{wikle2010}, in 1927, Yule used pendulum motion differential equations to suggest an autoregressive time series for sunspot observations 
and Hotelling used approximations of differential equations to develop models for population growth in the United States. The integration of mechanistic information in statistical models is now a fundamental component of several sub-disciplines of statistics; e.g., statistical meteorology  \citep{berliner1999bayesian}, ecological statistics \citep{hefley2017mechanism}, epidemiology \citep{leach2020linking}, glaciology \citep{gopalan2018bayesian}, statistical oceanography \citep{herbei2008gyres, milliff2011ocean}, pharmacokinetic and pharmacodynamic analysis in the development of pharmaceuticals \citep[e.g.,][]{gabrielsson2001pharmacokinetic}, to name a few.

With the advent of deep learning, the use of mechanistic information combined with deep neural network models has seen an exponential increase in applications across many disciplines. Such approaches, often known as Physics Informed Neural Networks \citep[PINNs,][]{Raissi2017}, are typically based on augmenting the neural network loss (e.g., mean squared error) with loss terms corresponding to the physical equations and boundary/initial conditions \citep[e.g.,][]{Karniadakis}.  Although this is novel in the context of neural network models, augmenting loss functions to include mechanistic information when fitting models to data is not new. For example, Sasaki pioneered the approach in a variational solution to the data assimilation problem in meteorology beginning in the 1950s \citep[e.g.,][]{sasaki1958objective,sasaki1970some}. Around the same time, control engineers were combining (linearized) dynamical equations within a filter that updated in the presence of new observations \citep[e.g.,][]{kalman1960new}. Additional approaches developed in engineering and statistics to ``fuse'' data, random processes, and deterministic models were developed starting in the 1990s and continue to this day \citep[e.g., see overviews in][]{berliner2003physical, cockayne2019bayesian, meng2020survey}. 

From a statistical perspective, the goal is not to develop a ``solver'' for a known mechanistic set of equations as is often of interest in the PINN literature \citep{Raissi2019}, but rather to recognize, in the G. Box sense, that all our mechanistic models are in some way ``wrong'' when applied to real world scenarios, and that we need to account for the uncertainty in the model specification, parameters, and data.  In this sense, the Bayesian hierarchical modeling (BHM) approach has long been seen as the natural framework to account for all of these sources of uncertainty in physics-motivated statistical modeling \citep[e.g., see][]{berliner1999bayesian,cressie2011statistics,hennig2015probabilistic}.  The Bayesian view has also increasingly been shared in the PINN community \citep[e.g.,][]{yang2021b,meng2022learning,huang2023posterior,psaros2023uncertainty}.  This has been an important development from a statistical perspective.  Yet, some components of the BHM modeling paradigm are often not explicitly represented in these implementations, and there is still very little consideration of the importance of specifying dependent random effects, which has been shown to be very useful in statistical mechanistic modeling of spatio-temporal processes \citep[e.g.,][]{wikle2001spatiotemporal,cressie2011statistics,chkrebtii2016bayesian}.

Our goal in this paper is to present a statistically-minded overview of mechanistically-informed modeling in the presence of data and uncertainty to provide context in our discussion of PINNs. Note, although we generally use the PINN acronym due to its ubiquitous use in the literature, we prefer the term ``mechanistically-informed'' to ``physics-informed'' because many mechanistic models for non-physical processes (e.g., invasive species in ecology, \citet{wikle2003hierarchical}) also follow this paradigm and it can be misleading to use the term ``physics'' for such processes.   We will illustrate how it is natural to consider a PINN within a traditional BHM as is commonly considered for dynamic spatio-temporal statistical models \citep[e.g., see][]{cressie2011statistics}; we call this a BHM-PINN.  We first provide  a brief historical overview of how mechanistic process information has been combined with data in various modeling paradigms. Throughout, we discuss issues of interest to statisticians, such as uncertainty in data, parameters, and process knowledge, as well as computational tractability. 

Section \ref{sec:ProbOut} discusses the general problem at hand and introduces the relevant notation. Then, in Section \ref{sec:EarlyLiterature} we present a brief review of mechanistically-informed modeling from the perspective of variational data assimilation and optimal interpolation, as well as statistical approaches, with a focus on Bayesian and hierarchical Bayesian implementations to account for uncertainty.  Section \ref{sec:4} describes PINNs and Bayesian-PINNs (B-PINNs) as well as connections to the aforementioned earlier literature. Section \ref{sec:BHM-PINN} discusses a how it is natural to include a B-PINN in a traditional BHM framework for dynamic spatio-temporal models, which we call a BHM-PINN.  
In Section \ref{sec:simul}, we apply the BHM-PINN framework to Poisson distributed spatio-temporal data where the latent process corresponds to a mechanistic PDE parameterized using Burgers' equation. We conclude with a brief overview and discussion of challenges in Section \ref{sec:Conclusion}.

\section{Problem Outline}
\label{sec:ProbOut}
Our interest in mechanistically-informed models here concerns problems common to spatio-temporal statistics where one has potentially incomplete and noisy observations of a univariate or multivariate process at some locations in space and time, and seeks to: (a) determine an initial condition (i.e., the true process values at specified spatial locations at some initial time), (b) predict (interpolate) incomplete observations at specified locations in space and time, or (c) forecast at specified spatial locations at times beyond which we have observations.  We generally assume that we have some {\it a priori} knowledge of the underlying spatio-temporal dynamics of the process, yet our knowledge is incomplete. That is, we may not know the parameters of the mechanistic model and/or the mechanistic model is an incomplete representation of the process. An incomplete representation of dynamics could be due to many circumstances -- for instance, due to approximations to the underlying partial differential equations, or in other instances due to missing mechanistic understanding. Next, we outline the notation to be used in this exposition.

\subsection{Process and Data Notation}\label{sec:notation}
Assume our interest is in a $K$-variate process at spatial locations $\bm{s} \in \mathcal{D}$ (in $\mathbb{R}^1, \mathbb{R}^2$, or $\mathbb{R}^3$) and times $t \in [0,T+\tau]$, denoted by
$
\bm{u}_t(\bm{s}) \equiv (u^{(1)}_t(\bm{s}),\ldots,u^{(K)}_t(\bm{s}))'$.
Ultimately, we are interested in this process at some fixed set of $n$ spatial locations, say $\mathcal{S} \equiv \{\bm{s}_1,\ldots,\bm{s}_n\}$ and some discrete set of $n_{\mathcal{T}}$ times $\mathcal{T} \subseteq \{0,1,\ldots,T+\tau\}$ (although, the approach presented here can accommodate continuous time).
If we stack these vectors, we can write the $Kn$-vector
$
\bm{u}_t \equiv (\bm{u}_t(\bm{s}_1)',\ldots,\bm{u}_t(\bm{s}_n)')'$.
For example, if the process is univariate then $K=1$ and $\bm{u}_t \equiv (u_t(\bm{s}_1),\ldots,u_t(\bm{s}_n))'$. More generally, observations could be available at different locations for each of the $K$ variables of interest, but we do not add that additional notational complexity here.

Assume we have observations of elements of the process corresponding to some or all of the times in $\mathcal{T}$ at $m_t$ spatial locations, which we denote by the $m_t$-vector $\bm{z}_t$. Some or all of these observation locations  may coincide with the spatial locations in $\mathcal{S}$, but may not, as discussed below.  Note, it may be the case that $m_t = 0$ for some times (i.e., there are no observations for certain times).

\subsection{ Mechanistic Model Notation}\label{sec:mechmodel}

Here we focus our attention on mechanistic models that take the form of a time-dependent partial differential equation (PDE), but the ideas presented here can be easily adapted to other forms of mechanistic models (e.g., ordinary differential equations, or agent-based models).  Assume we have a PDE that we believe is mechanistically informative to our latent process of interest, $\bm{u}_t(\bm{s})$. We denote this as
\begin{equation}
\frac{\partial \bm{v}_t(\bm{s})}{\partial{t}} = \mathcal{M}(\bm{v}_t(\bm{s}), \bm{\theta}_m),
\label{eq:PDE}
\end{equation}
where $\bm{v}_t(\bm{s})$ corresponds to the $K$-dimensional mechanistic process defined at $t \in [0,T+\tau]$ and $\bm{s} \in \mathcal{D}$ and $\mathcal{M} $ is an operator controlled by parameters $\bm{\theta}_m$. For example, we may believe that our latent process shows evidence of advective and diffusive dynamics, both of which can be represented through appropriate terms in $\mathcal{M}$.  Note, we are deliberately making a distinction between the true latent process of interest, $\bm{u}_t(\bm{s})$, and the deterministic mechanistic model representation, denoted by ${\bm{v}}_t(\bm{s})$ - this distinction will be made more clear when we discuss models in Section \ref{sec:EarlyLiterature} and \ref{sec:4}.  We also assume that the PDE has associated initial and boundary conditions, respectively,
\begingroup
\setlength{\abovedisplayskip}{3pt} 
\setlength{\belowdisplayskip}{3pt} 
\begin{align*}
\bm{v}_0(\bm{s}) &= \bm{g}_I(\bm{s}), \;\; \bm{s} \in \mathcal{D}, \\
\bm{B}(\bm{v}_t(\bm{s})) & =0, \;\; t \in [0,T+\tau], \bm{s} \in \mathcal{D}_B, 
\end{align*}
\endgroup
where $\bm{g}_I(\cdot)$ is some function defined at the initial time, 0, $\bm{B}(\cdot)$ is a function defined on the boundary $\mathcal{D}_B$. 

Our goal is to perform inference on $\bm{u}_t(\bm{s})$ at the $\mathcal{S}$ spatial locations and at times $t \in \mathcal{T}$ while taking into account the information present in $\mathcal{M}$ and given the data $\bm{Z} \equiv \{\bm{z}_1,\ldots,\bm{z}_T\}$. We can denote this as a posterior distribution (with some abuse of notation) as
$[\bm{u}_t: t \in \mathcal{T} | \bm{Z}, \mathcal{M}]$,
where we use the bracket notation ($[ \;\;]$) to represent a probability distribution, as is common in the Bayesian hierarchical modeling literature \citep[e.g.,][]{gelfand1990sampling,cressie2011statistics}; that is, $[a | b]$ corresponds to the distribution of $a$ given (conditional upon) $b$, $[b]$ corresponds to the marginal distribution of $b$, and $[a,b]$ corresponds to the joint distribution of $a$ and $b$. Typically, we are interested in learning the latent process at some initial time $\bm{u}_0$, interpolation at some time $t \in (0,T]$, $\bm{u}_t$, or forecasting at some time $T+\tau$,  $\bm{u}_{T+\tau}$.
This is generally termed the ``forward problem'' where $\mathcal{D}, \mathcal{D}_B, \bm{g}_I(\cdot), \bm{B}(\cdot)$, and $\bm{\theta}_m$ are known and one seeks to find $\bm{u}_0(\bm{s})$ and $[\bm{u}_t]_{t \in (0,T+\tau]}$ given the dynamics suggested by $\mathcal{M}$.
We may also be interested in learning parameters associated with the mechanistic PDE ($\bm{\theta}_m$) and various parameters associated with the underlying statistical or neural models used to represent the PDE. 
This is generally termed the ``inverse problem,'' where we may have partial observations of the process in the domain of interest. The primary goal of the methodology presented in this work is to infer the latent process $\{\bm{u}_t\}_{t \in [0,T+\tau]}$ on $\mathcal{D}$ as well as the $\bm{\theta}_m$ that satisfies $\mathcal{M}$ given observations $\bm{Z}$. 

In the exposition that follows, we will sometimes refer to the numerical solution of the PDE in the short-hand notation,
\begin{equation}\label{eq:NumPDE}
\bm{v}_{t+1} = \mathcal{M}(\bm{v}_t;\bm{\theta}_m),
\end{equation}
but recognize that there are PDE solvers that provide $\bm{v}_t$ for any and all times in $[0,T]$ that are not necessarily sequential in time (e.g., implicit solvers).

\section{Historical Perspective}
\label{sec:EarlyLiterature}
Fundamental advances in mechanistic-informed modeling in the presence of data originated with the operational need to assimilate observations with physical models of the atmosphere for weather forecasting (i.e., numerical solutions of simplified Navier-Stokes PDEs).  The main purpose of such ``data assimilation'' is to generate initial conditions for the numerical forecast models that are consistent with the dynamics of the system, but these initial conditions have incomplete observations in space (e.g., we cannot observe the atmosphere everywhere). We discuss the data assimilation approach in Section \ref{sec:sasaki}. The connection to more statistical approaches and further developments are discussed in Section \ref{sec:StatApproaches}.

\subsection{Variational Data Assimilation}
\label{sec:sasaki}

It was known shortly after L.F. Richardson's failed attempt at the first numerical weather forecast in the early 1920s \citep{richardson1922weather} that the numerical solution of PDEs with initial conditions given by data would quickly become unstable due to lack of consistency with the governing equations.  Thus, by the time that computers were actually capable of generating a realistic forecast in real time, it was known that initial condition fields of weather variables had to be smoothed in a way that respected dynamical constraints \citep{charney1950numerical,phillips1956general}. Two main approaches were developed in the 1950s and 1960s to do this, one deterministic based on the calculus of variations and one stochastic, based on what would come to be known as Gaussian processes (GPs). 
In this section, we briefly discuss these approaches as they provide the foundation for mechanistically-informed models for spatio-temporal processes in statistics. 

The initial idea in data assimilation, that one could minimize an objective function containing the sum of squared differences between the observations and the ``true'' initial field while simultaneously satisfying a constraint that incorporated the PDE dynamics, was presented in \citet{sasaki1958objective}. Although he formulated this in continuous space, practical uses of the methodology considered discrete space \citep[e.g.,][]{thompson1969reduction,sasaki1970some,stephens1970variational}. 

Following the notation in Section \ref{sec:notation}, say we have $m_t$-dimensional data vectors, $\bm{z}_t$, and numerical solutions of the deterministic PDE state process at time $t$ denoted by $\bm{v}_t$ given by (\ref{eq:NumPDE}). Here, without loss of generality, we assume $K=1$ (a univariate process) for notational simplicity. Critically, for deterministic systems, if one knows the initial state $\bm{v}_0$ then (\ref{eq:NumPDE}) determines the sequence of future states, $\{{\bm v}_1,{\bm v}_2,\ldots,{\bm v}_t,\ldots\}$.  In Sasaki's variational approach, one seeks an initial state ${\bm v}_0$ that combines the physics from the PDE and observations to obtain the initial condition. Importantly, in the context of weather forecasting, it is this estimated initial condition that is used to generate forecasts from the deterministic model, $\{ {\bm v}_t, t=1,\ldots \}$.

Sasaski's approach estimates the initial state, $\bm{v}_0$, by minimizing an objective function, $J_0(\bm{v}_0)$, subject to the constraint that the state evolution follows the numerical PDE solver exactly. That is, \citep[e.g., see][]{lewis2008sasaki} the optimization problem is to minimize
\begin{equation}
\label{eq:J0}
J_0(\bm{v}_0) = \frac{1}{2}\sum_{t=1}^T (\bm{z}_t - \bm{h}(\bm{v}_t))' \bm{R}_t^{-1} (\bm{z}_t - \bm{h}(\bm{v}_t)),
\end{equation}
subject to the constraint of (\ref{eq:NumPDE}) -- i.e., $\bm{v}_{t+1} = \mathcal{M}(\bm{v}_t;\bm{\theta}_m)$ for $t = 0, ..., T-1$; this constraint allows each of $\bm{v}_1,..., \bm{v}_T$ to be written as a function of $\bm{v}_0$ by iterating application of $\mathcal{M}(.;\bm{\theta}_m)$. Additionally, $\bm{h}(\cdot)$ is a (known) mapping function that takes the  $n$-dimensional deterministic model state vector $\bm{v}_t$ and maps to the $m_t$-dimensional observation vector at time $t$ (e.g., an interpolator if data are not coincident with model locations, or perhaps the identity if they are; this function can also take into account nonlinear transformations, see \citet{daley1993atmospheric}). In addition, $\bm{R}_t$ is the time-varying $m_t \times m_t$ observation error covariance matrix (representing measurement error and other sources of observational error), also typically assumed known. \citet{lewis2008sasaki} also describe how this constrained optimization problem can be solved via the method of Lagrange multipliers.  We emphasize that this approach is considered a ``strong'' constraint because the state solution must follow the deterministic model exactly \citep{sasaki1970some}.

\subsubsection{Operational Variational Data Assimilation (4DVar)}\label{sec:4dvar}

In practice, variational data assimilation evolved away from Lagrange multipliers to an unconstrained approach known as ``4DVar'' (``4D'' because it includes 3 dimensions in space and time). As with the Sasaki approach, one considers data in the interval $[1,T]$ and seeks the initial condition, $\bm{v}_0$. The observation component of the objective function is the same as (\ref{eq:J0}) above where again the $\bm{v}_t$ are functions of the initial condition $\bm{v}_0$ and obtained deterministically from the numerical solution of the PDE (\ref{eq:NumPDE}), which is assumed to be error free. 
In this approach, one then assumes that there is some background ``estimate'' of the initial state condition, say $\bm{v}_b$, such as a forecast from the previous assimilation cycle, or some historical average. Then, the background term of the objective function is given by,
\begin{equation}
J_{1b}(\bm{v}_0) = \frac{1}{2}(\bm{v}_0 - \bm{v}_b)' \bm{C}_{b_0}^{-1}(\bm{v}_0 - \bm{v}_b),
\label{eq:J1b}
\end{equation}
where $\bm{C}_{b_0}$ is the background error covariance matrix (a spatial covariance matrix for the errors between the true initial state and the previous forecast of the initial state). The optimization finds $\bm{v}_0$ that minimizes the sum of $J_0(\bm{v}_0) + J_{1b}(\bm{v}_0)$ (i.e., equation (\ref{eq:J0}) plus (\ref{eq:J1b})). In practice, the optimization is facilitated by using an adjoint method \citep[e.g.,][]{lewis1985use,errico1997adjoint}. 

\subsubsection{Weak Constraint 4DVar}
One can consider weak-constraint variational assimilation by incorporating an explicit model error term in the objective function \citep[e.g.,][via the method of representors]{bennett1992inverse}. For example, say we have the model error term, $\bm{\eta}_t$, as in
\begingroup
\setlength{\abovedisplayskip}{3pt} 
\setlength{\belowdisplayskip}{3pt} 
\begin{equation}
\bm{u}_t = \mathcal{M}(\bm{u}_{t-1};\bm{\theta}_m) + \bm{\eta}_t,
\label{eq:NumPDEerror}    
\end{equation}
\endgroup
where $\bm{\eta}_t \sim \text{N}(\bm{\mu}_{\eta}, \bm{Q})$, and the mean $\bm{\mu}_{\eta}$ vector and error covariance matrix $\bm{Q}$ are assumed known or estimated offline. In this section, we prefer to denote the process by $\bm{u}_t$ instead of ${\bm v}_t$ since $\bm{u}_t$ is not constrained to follow the numerical model exactly, as in the previous subsections.

Then, the extra term in the objective function in the 4DVar case is given by \citep[e.g.][]{laloyaux2020exploring},
\begin{equation}
J_2(\bm{\eta}) = \frac{1}{2} \sum_{t=1}^{T} (\bm{\eta}_t - \bm{\mu}_{\eta})' \bm{Q}^{-1} (\bm{\eta}_t - \bm{\mu}_{\eta}),
\label{eq:J2}
\end{equation}
where $\bm{\eta}_{t} \equiv \bm{u}_{t} - \mathcal{M}(\bm{u}_{t-1};\bm{\theta}_m)$ from (\ref{eq:NumPDEerror}) and $\bm{\eta} = \{\bm{\eta_1}, ..., \bm{\eta_{T}}\}$. Minimizing the loss function $J_0(\bm{u_0}, \bm{\eta}) + J_{1b}(\bm{u_0)} + J_2(\bm{\eta})$ (i.e., equation (\ref{eq:J0}) plus (\ref{eq:J1b}) plus (\ref{eq:J2})) in terms of $\bm{u}_0$ and $\bm{\eta}$ is constrained by the numerical model but can deviate from an exact solution of the model because of $\bm{\eta}$, resulting in a weak constraint. Note that $J_0$ is written as a function of both $\bm{u}_0$ and $\bm{\eta}$ because both are necessary to determine $\bm{u}_1, ..., \bm{u}_T$ as per (\ref{eq:NumPDEerror}).

\subsubsection{Optimal Interpolation}
\label{sec:3.2}
Around the same time that the variational approach to meteorological data assimilation was being developed, other researchers were developing statistical optimization approaches based on stochastic processes.  A brief history of these approaches is presented in \citet{lewis2008sasaki}. Although there is some debate about its origins, \citet{lewis2008sasaki} describe the first practical approach to come from \citet{eliassen1954provisional}. More formal development was given by \citet{gandin1963objective}. Gandin's approach is widely considered to have been equivalent to the formalization of kriging by Matheron \citep{matheron1963principles}, which is the essence of geostatistical modeling (``kriging'') in spatial statistics; see \citet{cressie1990origins} for a discussion of the origins of kriging.  In retrospect, all these approaches simply use Gaussian processes to perform interpolation (spatial prediction) that is optimal in the sense of minimizing the expected squared error loss (i.e., risk). 

In the context of meteorological data assimilation, one interpolates the residuals by removing a background field (e.g., the previous forecast) from observations. So, given an observation vector, $\bm{z}_t$, and a forecast from the previous time, $\bm{u}_b$ (where again, $b$ stands for ``background''; note, keeping with the notation used in this paper, when this forecast is from the numerical solution of a PDE, we could denote the background as ${\bm v}_b$), the kriging approach gives the latent state vector ${\bm u}_t$ from:
\begingroup
\setlength{\abovedisplayskip}{3pt} 
\setlength{\belowdisplayskip}{3pt} 
\begin{equation*}
{\bm{u}}_t = \bm{u}_b + \bm{K}_t(\bm{z}_t - \bm{H}_t\bm{u}_b),
\end{equation*}
\endgroup
where $\bm{H}_t$ is an $m_t \times n$ matrix that maps the state process to the observations at time $t$, and $\bm{K}_t$ is the $n \times m_t$ matrix of interpolation weights given by
\begingroup
\setlength{\abovedisplayskip}{3pt} 
\setlength{\belowdisplayskip}{3pt} 
\begin{equation*}
\bm{K}_t = \bm{C}_b\bm{H}_t'(\bm{H}_t\bm{C}_b\bm{H}_t' + \bm{R}_t)^{-1},
\end{equation*}
\endgroup
where $\bm{C}_b$ is the $n \times n$ spatial covariance matrix of the background (forecast) error. This matrix must be specified in terms of a covariance function, and because one may need to develop predictions from locations that are not observed, it comes from a covariance function associated with a Gaussian process \citep[e.g., see][]{cressie1993statistics, cressie2011statistics}. 

There are different ways that physical constraints are included in this framework. First, note that the forecast for time $t$, which we are denoting by $\bm{u}_b$ here, comes from a forecast given by the dynamic model (\ref{eq:NumPDE}) and the previous estimate of the state, say ${\bm{u}}_{t-1}$. Alternatively, one can specify covariance functions that satisfy dynamical constraints, which is especially useful in multivariate applications of this methodology \citep[e.g., see][]{daley1993atmospheric}.

Note, this formulation does not explicity account for multiple observations over times; i.e., there is no dependence specified across time. In addition, this optimal interpolation approach for spatial fields can also be formulated from a variational perspective as in Section \ref{sec:4dvar}, known as ``3DVar'' in the data assimilation literature. Spatial optimal interpolation and 3DVar are equivalent if the observation mapping function is linear, i.e., $\bm{h}(\bm{u}_t) = \bm{H}_t\bm{u}_t$ \citep[e.g.,][]{lewis2008sasaki}. 

\subsubsection{Connections to Filtering and Bayesian Estimation}

About the same time that Sasaki was developing the variational approach to combining data and mechanistic model information, control engineers were working on the same problem from an online filtering perspective. The classic example of this is the Kalman filter \citep[][]{kalman1960new}, which has become ubiquitous in control engineering and statistical state-space time-series and spatio-temporal analysis \citep[e.g., see][]{shumway2000time,wikle2007bayesian}. In the control engineering case, one typically assumes they have an initial condition and initial variance/covariance matrix and proceeds to generate a forecast for the next time; this forecast is then updated as new observations come online, and this updated state estimate is used to generate another forecast, etc.  The filtering equations explicitly account for model error, as in the weak constraint 4DVar case. 

It has long been known that there are connections between the variational optimization approach in data assimilation and Bayesian inference \citep[e.g.,][]{lorenc1986analysis,talagrand1997assimilation,wikle2007bayesian}. This is not surprising to statisticians, as the maximum {\it a posteriori} (MAP) estimator in a Bayesian model is equivalent to a regularized optimization. That is, say we have data $\bm{D}$ and parameters $\bm{\theta}$, then the MAP estimator is given by maximizing the numerator in Bayes formula with respect to $\bm{\theta}$
\begingroup
\setlength{\abovedisplayskip}{3pt} 
\setlength{\belowdisplayskip}{3pt} 
\begin{eqnarray*}
\hat{\bm{\theta}}_{MAP} & = & \underset{
\bm{\theta}}{\arg\max} ([\bm{D} | \bm{\theta}][\bm{\theta}])\\
&  = & \underset{
\bm{\theta}}{\arg\max} (\log [\bm{D} | \bm{\theta}] + \log[\bm{\theta}]),
\end{eqnarray*}
\endgroup
which of course is equivalent to minimizing $(-\log [\bm{D} | \bm{\theta}] - \log[\bm{\theta}])$. The connection is then obvious in that the first term (Bayesian likelihood) corresponds to a loss function such as (\ref{eq:J0}) and the second term (Bayesian prior) corresponds to a regularization term such as (\ref{eq:J1b}). A classic example in statistics is ridge regression, where a least squares regression is regularized by an $\ell_2$ penalty on the parameters, which is equivalent to a multivariate normal prior on the parameters. Typically, assuming one has a reasonably strong belief that their models are correct, the Bayesian approach is often preferred because one gets natural uncertainty quantification. However, in the context of high-dimensional nonlinear data and process models that are used in operational data assimilation, the variational approach has been an effective and reasonably efficient assimilation approach where fully Bayesian estimation would be computationally prohibitive. 

Similarly, the sequential estimation in the Kalman filter procedure is also naturally considered from a sequential Bayesian perspective  \citep[e.g.,][]{talagrand1997assimilation,wikle2007bayesian}. That is, the forecast distribution for time $t+1$ can be considered as the ``prior'' distribution for updating the latent state for time $t+1$ when new data become available for that time. For linear, normal data and process models, these update equations are available in closed form and are equivalent to the variational approach, but when these distributions are non-normal or the dynamics are nonlinear, then sequential Monte Carlo approaches such as particle filtering or ensemble Kalman filtering must be used \citep[e.g., see the tutorial in][]{katzfuss2016understanding}. 

One of the limitations of considering the equivalence of the variational and Bayesian perspective is that it assumes the model parameters (and error variances/covariances) are known (or they are somehow estimated off-line) and that one's models are correct.  In the case of online sequential filtering, one can augment the state vector with parameters and update them along with the state process. Alternatively, in variational approaches it is certainly possible to add another term to the objective function  that accounts for parameter uncertainty. This is equivalent to assigning a prior distribution to the parameters. When practical, it is often more flexible to specify prior distributions explicitly in case non-normal priors are warranted. The development of mechanistically-motivated spatio-temporal dynamic models (described in the next section) evolved from this hierarchical Bayesian perspective.

\subsection{Statistical Approaches for Adding Mechanistic Information} \label{sec:StatApproaches}

There have been several approaches to combine mechanistic information in statistical spatial and spatio-temporal models. For example, the development of spatial and spatio-temporal covariance functions from analytical or numerical solutions to PDEs has long been considered an efficient way of incorporating mechanistic information in spatial and spatio-temporal models. PDE-derived covariance functions have been used for spatial prediction, such as for kriging models \citep[e.g.,][]{heine1955models, whittle1954stationary}, and have recently become quite popular with the stochastic-PDE (SPDE) approach \citep[e.g.,][]{lindgren2011explicit, lindgren2022spde}.  Indeed, as mentioned in Section \ref{sec:3.2}, this idea of using PDE-derived covariance functions for optimal interpolation has also been used in the early data assimilation literature \citep[e.g.,][]{daley1993atmospheric}. Generally, this approach is  useful for developing covariance functions, but is quite limited in the complexity of the underlying mechanistic model that can be effectively considered \citep[see the overview in ][]{cressie2011statistics}. 

As an alternative, in the functional data and non-parametric regression analysis literature, it has long been common to use function derivatives to regularize a regression model such as in smoothing splines for spatial data \citep[e.g.,][]{wahba1990spline, eubank1999nonparametric,ramsay2002spline}.  Over the last 10 years or so, this has been extended to the physics-informed regularization case in spatial and spatio-temporal regression \citep[e.g., see the review in ][]{sangalli2021spatial}.  These methods are very much in the spirit of the \citet{sasaki1958objective} approach described in Section \ref{sec:sasaki}, where one  adds a PDE regularization term to the loss function that compares the non-parametric model of interest to the data.  As described in \citet{tomasetto2024modeling}, point estimates of parameters can be estimated in this framework. These approaches can be quite effective for certain problems, and in fundamental ways the Physics-Informed Neural Network (PINN) approach described below in Section \ref{sec:pinn} can be thought of as a generalization of this approach with more flexible semi-parametric (deep neural) models and the use of automatic differentiation to allow more complex PDE regularization.  Bayesian-PINNs (described in Section \ref{sec:B-PINN}), mechanistically-motivated statistical models (described in Section \ref{sec:MechMotMods}), and  probabilistic numerical methods (described in Section \ref{sec:PNM}) operate similarly, but can accommodate more principled uncertainty quantification across all phases of model development via a hierarchical Bayesian framework. Given that uncertainty quantification is a major focus for modern spatio-temporal modeling applications, we describe the statistical approaches in more detail in the following subsections, and PINNs and B-PINNs in Section \ref{sec:4}.

\subsubsection{Mechanistically-Motivated (Physical-Statistical) Models }\label{sec:MechMotMods}

Before introducing the mechanistically-motivated framework, we first briefly review Bayesian hierarchical models for spatio-temporal processes. Adding additional sources of data and accounting for uncertainty in observation model parameters and state-process model parameters is often needed when modeling complex spatio-temporal data. In some cases, one can gain much more expressiveness in their models if they add dependence structure (time, space) to the parameters as well by considering them to be spatial or temporal random processes and/or to be related to time or space-varying covariates. Bayesian hierarchical models are natural extensions of the Bayesian paradigm that allow for more complete accommodation of uncertainty and this additional model flexibility.

The Bayesian hierarchical (deep) modeling paradigm for spatio-temporal data \citep[e.g.,][]{berliner1996hierarchical, cressie2011statistics} considers breaking the joint distribution of data, process, and parameters into a series of conditional models (distributions). In that context, the posterior distribution is then given by the following general decomposition
\begingroup
\setlength{\abovedisplayskip}{3pt} 
\setlength{\belowdisplayskip}{3pt} 
\begin{eqnarray*}
[process, \; parameters \;  | \; data ] & \propto & [data\; |\; process, data \; parameters] \\
 & \;\;\; &  \times \;\;\; [process\; | \; process \; parameters]\\
 & \;\;\; &  \times \;\;\; [parameters].
\end{eqnarray*}
\endgroup
Importantly, each of these stages can have sub-stages as described below in Section \ref{sec:MechMotMods}.  

In the context of the dynamic spatio-temporal models (DSTMs) of interest here, the basic BHM might take the following form \citep[e.g., see][]{cressie2011statistics}.
\begin{align*}
\text{Data Model:} & \;\;\;
\bm{z}_t = \bm{h}_t(\bm{u}_t) + \bm{\epsilon}_t,  \;\; \bm{\epsilon}_t \overset{ind.}{\sim} \; \text{N}(\bm{0},\bm{R}_t(\bm{\theta}_z)),
\\
\text{Process (Dynamic) Model:} & \;\;\;
\bm{u}_t = \mathcal{M}(\bm{u}_{t-1},\bm{B}(\bm{u}_t;\bm{\theta}_b);\bm{\theta}_m) + \bm{\eta}_t, \;\; \bm{\eta}_t \overset{ind.}{\sim} \; \text{N}(\bm{0},\bm{Q}_t(\bm{\theta}_q)),\\
\text{Process Initial Value:} & \;\;\; \bm{u}_0 \sim \text{N}(\bm{a},\bm{\Sigma}_0) \\ 
\text{Parameter Models:} & \;\;\;
[\bm{\theta}_z, \bm{\theta}_m, \bm{a}, \bm{\Sigma}_0, \bm{\theta}_b, \bm{\theta}_q],
\end{align*}
where $\bm{h}_t( )$ is a mapping function between the latent process and observations, $\text{N}(\bm{\mu},\bm{\Sigma})$ is a multivariate normal distribution with mean ${\bm \mu}$ and covariance matrix ${\bm \Sigma}$, and $\bm{B}(\bm{u}_t;\bm{\theta}_b)$ represents a boundary condition and $\bm{u}_0$ is the initial condition, both of which depend on parameters that are given prior distributions in general. The unknown parameters are given by ${\bm \theta}_z, {\bm \theta}_m,  {\bm a}, {\bm \Sigma}_0, {\bm \theta}_b, {\bm \theta}_q,$. As mentioned above, the parameter models can also have dependence (in space and/or time) or depend on other covariates, which gives this model a great amount of generalizability and can make it much ``deeper'' in terms of having more levels of conditioning (e.g., see {\it Process} description below).

{\bf Including Mechanistic Information:} The BHM framework naturally allows one to include mechanistic information in each model layer. Here, we briefly describe how mechanistic information can be incorporated into the data, process, and parameter levels.

{\it Data:}  A strength of the BHM approach is that one can easily specify multiple data sources conditioned on the latent state process, e.g., $[Z_1, Z_2 | u] = [Z_1 | u][Z_2 | u]$ (assuming conditional, although this assumption can be relaxed).  In statistics, a common {\it ``data fusion''} approach since the late 1990s has been to combine {\it in situ} data, such as satellite observations ($Z_1$) and mechanistic model outputs such as weather reanalysis fields ($Z_2$), to incorporate mechanistic information \citep[e.g.,][]{berliner1999bayesian,wikle2001spatiotemporal,berliner2003physical,fuentes2005model,wikle2005combining}. This is a type of weak constraint mechanistic modeling that has the advantage of being able to account for additive errors (to facilitate uncertainty quantification). 

{\it Process:} As with the data assimilation literature, one can specify $\mathcal{M}(\cdot)$ to be a known mechanistic model. Alternatively, in the physical-statistical (mechanistically-motivated) DSTM paradigm, one can specify statistical models for the transition operator, say $\bm{M}(\cdot;\bm{\theta}_m)$, that are ``motivated'' by mechanistic processes, but are conditioned on parameters $\bm{\theta}_m$ that typically have spatial or temporal structure (e.g. spatially-varying diffusion parameters); see \citet{wikle2010} and \citet{cressie2011statistics} for numerous examples. For example, one may discretize (or write in spectral/finite-element space) a reaction-diffusion PDE or integro-difference equation and condition this on the diffusion/growth parameters that may vary with space as a GP, or in terms of covariates, which are specified at the next level of the hierarchy \citep[e.g.,][]{wikle2003hierarchical,xu2005kernel,wikle2007bayesian,hooten2008}.

{\it Parameters:} In some cases, mechanistic information can be imparted to the parameters (say, $\bm{\theta}_m$) through the prior specification; e.g., via regularization or informed by mechanistic information. For example, \citet{wikle2003hierarchical} allowed covariates and a latent Gaussian process to model the spatially-varying diffusion rate in reaction-diffusion model for spread of an invasive species and \citet{wikle2001spatiotemporal} included a known turbulent scaling relationship as a prior on the variances of tropical surface winds to add realistic structure. Alternatively, \citet{leeds2014emulator} used pre-training/meta-learning of prior distributions with model output obtained from a coupled ocean-ecosystem model to inform a quadratic nonlinear mechanistically-motivated statistical model for data assimilation of ocean color measurements.

As a concrete example illustrating the process and parameter stages of a mechanistically motivated DSTM, consider the latent dynamic process $v$ in one-dimensional space, which is assumed to approximately follow the advection–diffusion equation 
\begingroup
\setlength{\abovedisplayskip}{3pt} 
\setlength{\belowdisplayskip}{3pt} 
\begin{equation}
\frac{\partial v}{\partial t} =  a \frac{\partial v}{\partial s} +  \lambda \frac{\partial^2 v}{\partial s^2}
\label{eqn:advecdiff}.
\end{equation}
\endgroup
As shown in \citet{wikle2007bayesian}, a simple finite difference numerical solution suggests the following difference equation
\begingroup
\setlength{\abovedisplayskip}{3pt} 
\setlength{\belowdisplayskip}{3pt} 
\begin{equation}
v_t(s) = \theta_1 v_{t-\Delta_t}(s) + \theta_2 v_{t-\Delta_t}(s + \Delta_s) + \theta_3 v_{t-\Delta_t}(s - \Delta_s), \nonumber
\end{equation}
\endgroup
where the coefficients $\theta_i$ are functions of the spatial discretization, $\Delta_s$, time discretization, $\Delta_t$, and advection and diffusion parameters, $a$ and $\lambda$, respectively. Denoting the discretized process at $n$ spatial locations by the vector ${\bm v}_t = [v_t(s_1),\ldots,v_t(s_n)]'$, and the boundary process by ${\bm v}_t^b = [v_t(s_0),v_t(s_{n+1})]'$, we can write the vectorized difference equation for the solution of the PDE as ${\bm v}_t = {\bm M} {\bm v}_{t-\Delta_t} + {\bm M}^b {\bm v}^b_{t-\Delta_t}$,
where ${\bm M}$ and ${\bm M}^b$ are functions of $\theta_0,\theta_1,\theta_2$; e.g., ${\bm M}$ is a tridiagonal matrix (e.g., with $\theta_0$ on the main diagonal and $\theta_1, \theta_3$ on the off-diagonals; e.g., see \citet{wikle2007bayesian}). Now, a mechanistically motivated model for the latent process of interest, say ${\bm u}_t$, could be written as ${\bm u}_t = {\bm M}_u {\bm u}_{t-1} + {\bm \eta}_t$, where ${\bm M}_u$ might be a tridiagonal matrix, but where the main diagonal is spatially varying, e.g., ${\bm \theta}_0 \sim \text{N}({\bm X} {\bm \beta}, {\bm \Sigma}_\theta)$, where ${\bm \theta}_0$ is an $n \times 1$ vector and $\bm X$ represents covariates and $\bm \Sigma_\theta$ imparts spatial dependence. With a data model that relates observations to the latent process, ${\bm u}_t$, this corresponds to a simple version of the BHM DSTM described above. The aformentioned references contain specific examples and extensions of such a mechanistically motivated model. For example, if desired, one can include hard constraints (i.e., ``mass balance'') in the physical-statistical framework \citep[e.g.][]{berliner2008modeling, zhuang2014bayesian,gopalan2018bayesian}.

\subsubsection{Probabilistic Numerical Methods}
\label{sec:PNM}
Over the last decade, there has been growing interest in modernizing the ideas associated with treating differential equation solutions as random processes \citep[e.g.,][]{larkin1972gaussian,diaconis1988bayesian}. These ideas, which come under a broad set of methods called {\it probabilistic numerical methods} (PNMs), have provided a comprehensive toolkit to consider uncertainty quantification in the solution of differential equations \citep[e.g., see][and references therein]{hennig2015probabilistic,chkrebtii2016bayesian,cockayne2019bayesian}. These methods have the advantage of being rigorous in their convergence guarantees and are straightforward to implement.

Many PNM methods for solving PDEs with uncertainty assume the solution has a GP prior. Then observations of boundary conditions and forcing terms of the PDE at a finite set of (collocation) points act as data that allow for a posterior distribution over the PDE solution. For a simple illustration, in the space-only case, we might have the following GP prior:
\begingroup
\setlength{\abovedisplayskip}{3pt} 
\setlength{\belowdisplayskip}{3pt} 
\begin{equation*}
u_t(s) \sim GP(m(s),k(s,s';\theta_k)),
\end{equation*}
\endgroup
where $m(s)$ is a mean function and $k(s,s';\theta_k)$ is a covariance function between two spatial locations $s$ and $s'$, defined by parameters $\theta_k$. Thus, this prior accommodates belief about the inherent smoothness of the PDE solution.  In the case of a linear map (say, $\mathcal{M}(u(s)) = f(s)$), the implied GP prior for the PDE solution is
\begingroup
\setlength{\abovedisplayskip}{3pt} 
\setlength{\belowdisplayskip}{3pt} 
\begin{equation*}
    \mathcal{M}(u(\bm{s})) \sim GP(\mathcal{M}(m(s)), \mathcal{M}{\mathcal M}'(k(s,s';\theta_k))),
\end{equation*}
\endgroup
where $\mathcal{M}{\mathcal M}'(k(s,s'; \theta_k))$ is the quadratic form associated with the linear operation on the prior covariance function (i.e., the kernel derived based on the fact that  $\bm{A}\bm{\Sigma}\bm{A}^{T}$ is the variance-covariance matrix of random vector $\bm{A} \bm{x}$ where the variance-covariance matrix of $\bm{x}$ is $\bm{\Sigma}$). One then treats forcing terms as data, e.g.,
\begingroup
\setlength{\abovedisplayskip}{3pt} 
\setlength{\belowdisplayskip}{3pt} 
\begin{equation*}
    f(s_i) = \mathcal{M}(u(s_i)) + \epsilon_i, \;\; \epsilon_i \sim \text{N}(0,\sigma^2_{\epsilon} ),
\end{equation*}
\endgroup
where $\sigma^2_{\epsilon}$ controls the fidelity of the PDE solution. Given the prior and data models are normal, the posterior distribution of $u(s)$ given $f(s_1),\ldots,f(s_m)$ are available in closed form via a normal distribution. In practice, one includes a boundary condition process and, for time-varying PDEs, an initial condition. In addition, one typically chooses the collocation points at which to obtain ``observations'' $f(s_i)$ to facilitate computation. This approach is often embedded in a BHM with real world observations to learn the parameters of the underlying PDE (i.e., the inverse problem). 

The strength of the PNM approach is its principled uncertainty quantification, which comes from the numerical approximation (e.g., the $f(s_i)$ observations) and, in the case of inverse problems, real-world observations. The inherent regularization of the Bayesian prior also provides some robustness to ill-posedness. Perhaps most importantly, PNM's strong connections to traditional numerical methods have been established for special cases (e.g., linear PDEs and GP priors).  That said, these methods are much more difficult to implement for nonlinear PDEs, do not scale well for very high-dimensional problems, and have not been used as much in deep BHMs with multiple sources of process uncertainty, as in the physical-statistical paradigm. However, the Bayesian PINN framework described in the next section might be considered a neural network based PNM.

\section{Physics-Informed Neural Networks (PINNs)}
\label{sec:4}
In the spirit of the Sasaki's variational approach described above in Section \ref{sec:sasaki}, consider fitting a deep neural network (NN) but constraining the solution to follow a PDE. This is the essence of PINNs \citep[][]{Raissi2017, Raissi2017a, Raissi2018d, Raissi2019, Raissi2020}. Such approaches can be used in the context of fitting a PDE in the presence of data (as in data assimilation), or solving a PDE explicitly (presumably for the sake of efficiency). We are interested here in the situation where we have data corresponding to the latent process of interest, initial conditions, or boundary conditions. 
In this section, we provide a brief overview of the PINN approach and its Bayesian counterpart (B-PINN).

\subsection{PINN Formulation}
\label{sec:pinn}
PINNs, first proposed by \citet{Raissi2017,Raissi2017a}, mainly consist of two components: (1) a neural network for prediction and (2) a physically-inspired equation (e.g., PDE) to impose constraints in the loss function. We denote the deep neural network solution to the PDE at time $t$ and location  $\bm{s}$ as $\bm{v}_{NN,t}(\bm{s};\bm{\theta}_W)$, where $\bm{\theta}_W$ represents the neural network parameters (weights and biases). Note, $\bm{v}_{NN,t}$ is an approximation of the PDE solution ${\bm v}_t$.  Indeed, since this is not a perfect representation it is best to make the distinction as follows by defining a PDE ``residual'' (as in the PNM approach described in Section \ref{sec:PNM}). Let $\bm{\theta}_m$ correspond to PDE parameters as in (\ref{eq:PDE}), and write 
\begin{equation}
\bm{r}_t(\bm{s}; \bm{\theta}_W,\bm{\theta}_m) \equiv  \frac{\partial {\bm{v}_{NN}(\bm{\theta}_W)}}{\partial{t}} -\mathcal{M}({\bm{v}_{NN}(\bm{\theta}_W)}; 
\bm{\theta}_m),
\label{eq:PDEresidual}
\end{equation}
for the ``residuals'' of the PDE evaluated at time $t$ and location $\bm{s}$ (note, in some cases where there is an exogenous forcing term in the PDE, this ``residual'' could be thought of as a representation of the forcing). In practice, as in the PNM approach, one considers $N_c$ so-called {\it collocation points} at which to evaluate the PDE, say  $\{\bm{s}_c^i,t_c^i\}_{i=1}^{N_c}$. Note, there can be different numbers of spatial locations at different times, so the $N_c$ collocation points include the complete set of space/time locations to train the neural net on the PDE (although we typically are interested in the solution at other locations in time and space). The loss function associated with the PDE is then given by the mean squared error (MSE),
\begin{equation}
J_{P}(\bm{\theta}_W) = \frac{1}{N_c} \sum_{i=1}^{N_c} \|\bm{r}_{t_c^i}(\bm{s}_c^i;\bm{\theta}_W)\|_2^2,
\label{eq:JP}
\end{equation}
where $\| \; \|_2$ represents the $\ell_2$ norm and, for now, we are assuming the PDE parameters $\bm{\theta}_m$ are known, and so exclude them notationally.
We typically would also include similar terms for the initial and boundary conditions, i.e., where there are $N_{IC}$ initial condition collocation points and $N_{BC}$ boundary condition collocation points. Denote these initial and boundary collocation points as $\{\bm{s}_{IC}^i\}_{i=1}^{N_{IC}}$ and  $\{\bm{s}_{BC}^i,t_{BC}^i\}_{i=1}^{N_{IC}}$, respectively; then the corresponding loss functions can be written as:
$$
J_{IC}(\bm{\theta}_W) = \frac{1}{N_{IC}} \sum_{i=1}^{N_{IC}} \|{\bm{v}}_{NN,0}(\bm{s}_{IC}^i;\bm{\theta}_W) - \bm{g}_I(\bm{s}_{IC}^i)\|_2^2,
$$
$$
J_{BC}(\bm{\theta}_W) = \frac{1}{N_{BC}} \sum_{i=1}^{N_{BC}} \|\bm{B}({\bm{v}}_{NN,t_{BC}^i}(\bm{s}_{BC}^i;\bm{\theta}_W))\|_2^2.
$$

If one is simply interested in finding the solution of the PDE given the initial and boundary conditions and PDE parameters $\bm{\theta}_m$, then one finds the $\bm{\theta}_W$ that minimize the combined objective function,
\begingroup
\setlength{\abovedisplayskip}{3pt} 
\setlength{\belowdisplayskip}{3pt} 
\begin{equation*}
J(\bm{\theta}_W) = a_{IC} J_{IC}(\bm{\theta}_W) + a_{BC} J_{BC}(\bm{\theta}_W) + a_{P} J_{P}(\bm{\theta}_W),
\end{equation*}
\endgroup
where the specified non-negative constants $\{a_{IC},a_{BC},a_{P}\}$ allow one to focus more on different components of the solution if desired. The solution in this case does not require measurements of the process besides initial and boundary condition training data, so the PINN serves as an alternative to the traditional numerical solver (e.g., finite difference, finite element or spectral scheme). 
In this context, note a well-known limitation of the PINN approach is it can have difficulty in simulating multi-scale, chaotic, or turbulent behavior because the explicit forward (``causal'') nature of time is not considered \citep[e.g.,][]{wang2022respecting}. 
This limitation can be mediated in a number of ways, for example, by adding a time weighting function in the loss function (\ref{eq:JP}). Regardless, some recent comparison studies have shown that traditional finite element solutions of PDEs work better and are as or more efficient than PINNs for most problems \citep[e.g.,][]{grossmann2024can}. 

\subsubsection{PINNs with Observational Data}
\label{sec:PINNwithObs}
As mentioned previously, in most cases in statistical applications of mechanistic-informed models with neural networks, our interest is to fit a deep neural network model with PDE constraints to data, which in the PINN literature is considered to be ``data-driven discovery'' of the PDE, by which they mean to learn the PDE parameters, $\bm{\theta}_m$ (i.e., an inverse problem). In this case, a data component is added to the objective function as in the variational approach and the PDE parameters can be estimated jointly with neural network parameters by minimizing the associated loss; the loss function $J_{P}$ in (\ref{eq:JP}) is a function of $\bm{\theta_m}$ in addition to $\bm{\theta_W}$ due to the dependence of $\bm{r}_t(\cdot)$ on $\bm{\theta_m}$.  Moreover, additional data observations of the process can be used; i.e., given observations, $\{\bm{z}_{t^i}(\bm{s}^i):i=1,\ldots,N\}$,
\begin{equation}
J_D(\bm{\theta}_W) = \frac{1}{N}\sum_{i=1}^N \|\bm{z}_{t^i}(\bm{s}^i) - \bm{v}_{NN,t^i}(\bm{s}^i;\bm{\theta}_W)\|_2^2.
\label{eq:JD}
\end{equation}
Note, in the original presentation of data-driven discovery, there is no measurement error associated with the observation in (\ref{eq:JD}).  The solution proceeds by optimizing the weighted combination 
\begingroup
\setlength{\abovedisplayskip}{3pt} 
\setlength{\belowdisplayskip}{3pt} 
\begin{equation*}
 J(\bm{\theta}_W, \bm{\theta}_m)  = a_{IC} J_{IC}(\bm{\theta}_W) + a_{BC} J_{BC}(\bm{\theta}_W) + a_{P} J_{P}(\bm{\theta}_W,\bm{\theta}_m) + a_D J_D(\bm{\theta}_W),
\end{equation*}
\endgroup
where we now explicitly note that $J_P$ depends on the PDE parameters $\bm{\theta}_m$. After this optimization, one has learned the PDE parameters and the associated neural network can be used to obtain an estimate of the latent process of interest, $\bm{u}_t(\bm{s})$, at desired locations in space and time. 

Importantly, ``data-driven discovery of PDEs'' does not mean finding a functional form of PDEs in this case, but rather indicates estimating parameters within PDEs. The use of statistical models and deep neural networks to learn the functional form of mechanistic models is a growing area of research, but beyond the scope of this article. For recent overviews, see \citet{north2023review}, \citet{sahimi2024physics}, and \citet{north2025bayesian}.

\subsection{Bayesian PINNs}
\label{sec:B-PINN}

The PINN approach described above does not explicitly account for uncertainty in the data, model, or parameters. It is quite straightforward to account for measurement error as in the variational approach presented earlier; we can modify the data portion of the loss function (\ref{eq:JD}) by assuming that the measurement errors are normal.  For example, we might assume that we have observation models of a univariate process (although this can be generalized to a multivariate process)
$
z_{t^i}(\bm{s}^i) = v_{NN,t^i}(\bm{s}^i;\bm{\theta}_W) + \epsilon_{t^i}(\bm{s}^i),
$
where $\epsilon_{t^i}(\bm{s}^i) \sim \text{N}(0,\sigma^2_z)$. Similarly, in principle we could have observations of $r_t(\bm{s})$ (the PDE residual term), boundary condition $B(\cdot)$ and initial conditions $g_I(\cdot)$ that also have independent normal errors and associated error variances. 
All these distributions are conditioned on the solution of the PDE and PDE model parameters, which in the PINN context are completely determined given the NN and PDE model parameters, $\bm{\theta}_W, \bm{\theta}_m$. Note, in principle, one could have dependent errors as described in the variational and physical-statistical approaches above. 

It is then natural, following from the Bayesian NN literature \citep[e.g.,][]{neal2012bayesian}, to specify prior distributions on the NN weight and bias parameters, $\bm{\theta}_W$; known as a B-PINN \citep[e.g.,][]{yang2021b}. That is, assume we generically have conditional data distributions for the process, PDE residual, boundary and initial conditions given by $[D_z | \sigma^2_z, \bm{\theta}_W]$, $[D_r | \sigma^2_r, \bm{\theta}_W, \bm{\theta}_m]$, $[D_{bc} | \sigma^2_{bc},\bm{\theta}_W]$ and $[D_{ic} | \sigma^2_{ic},\bm{\theta}_W]$, respectively. Note, we use $D_z$ to represent the observations $\{z_{t^i}(\bm{s}^i)\}$ described above to provide a more general and consistent notation for the observations. Each of these terms is implicitly conditioned on $v_{NN,t}$, but as mentioned above, since the NN is completely determined conditioned on its weights and biases, it is not necessary to write this explicitly. In addition,  as in the case of PNMs, unless we have a forcing term in the PDE, we assume implicitly that ``observations'' of the residuals are equal to zero on average -- that is, conditioned on the parameters, the relationship in (\ref{eq:PDE}) holds perfectly up to some arbitrary white noise (represented by $\sigma^2_r$). This has the effect of including a term equivalent to (\ref{eq:JP}) in the un-normalized posterior representation (\ref{eq:post}) below. That is, we can either write this as a ``pseudo data model'' with known zero observations or as a regularization prior term. For the latter, we might write the prior distribution as:
\begin{equation}
\frac{\partial {v_{NN}(\bm{\theta}_W)}}{\partial{t}} \sim \text{N}(\mathcal{M}({v_{NN}(\bm{\theta}_W)}; 
\bm{\theta}_m),\sigma^2_r),
\end{equation}
which corresponds to a regularization term and discrepancy parameter, $\sigma^2_r$, that effectively specifies {\it a priori} belief in the representativeness of the PDE for this process.

Assuming conditional independence of the different data types, one can express the conditional data distribution as
\begingroup
\setlength{\abovedisplayskip}{3pt} 
\setlength{\belowdisplayskip}{3pt} 
$$
[D_z,D_r,D_{bc}, D_{ic} | \bm{\theta}_W, \bm{\theta}_m, \bm{\theta}_D] = [D_z | \sigma^2_z, \bm{\theta}_W][D_r | \sigma^2_r, \bm{\theta}_W, \bm{\theta}_m][D_{bc} | \sigma^2_{bc},\bm{\theta}_W][D_{ic} | \sigma^2_{ic},\bm{\theta}_W],
$$
\endgroup
where $\bm{\theta}_D \equiv (\sigma^2_z,\sigma^2_r,\sigma^2_{bc},\sigma^2_{ic})'$. The parameter $\bm{\theta_m}$ is only included for the conditional distribution of $D_r$ because PDE residuals are the only quantities that depend on the physics parameters; the other quantities' conditional distributions are only dependent on the neural network, and consequently only $\bm{\theta}_W$. Now, assume we have prior distributions $[\bm{\theta}_W]$, $[\bm{\theta}_m]$, and $[\bm{\theta}_D]$. Then, Bayes' rule gives
\begingroup
\setlength{\abovedisplayskip}{3pt} 
\setlength{\belowdisplayskip}{3pt} 
\begin{equation}
[\bm{\theta}_W, \bm{\theta}_m, \bm{\theta}_D | D] \propto [D | \bm{\theta}_W, \bm{\theta}_m, \bm{\theta}_D ][\bm{\theta}_W][\bm{\theta}_m][\bm{\theta}_D],
\label{eq:post}
\end{equation}
\endgroup
where $D \equiv \{D_z, D_r, D_{bc},D_{ic}\}$.
The intractable posterior (\ref{eq:post}) is typically sampled from using MCMC methods (e.g., Hamiltonian Monte Carlo, Langevin Dynamics), variational methods, or ensemble methods \citep[see the comprehensive review in][]{psaros2023uncertainty}. 

A significant challenge with Bayesian NNs (and, hence, B-PINNs) is the specification of reasonable prior distributions for the NN parameters $\bm{\theta}_W$. In general, such parameters are not identifiable, which can make MCMC sampling challenging without extra constraints \citep[e.g.,][]{mcdermott2019bayesian}, the specification of informative priors (e.g., through empirical Bayes or meta-learning; e.g., \citet[][]{meng2022learning}), or special classes of priors \citep[e.g.,][]{meng2022learning,sell2023trace,castillo2024posterior}. 

Note that the B-PINN framework presented here does not include an explicit term for model error, although $\sigma^2_r$ does account for minor discrepancies in the PDE. But, this does not respect the fact that the PDE dynamics are not usually sufficient to model the real world process. In other words, the real-world dynamical process is latent and is motivated by the PDE, but is typically more complicated. This is exactly the motivation for denoting a distinction between the true latent process, $u$, and the PDE solution $v$ in Section \ref{sec:mechmodel} and the NN solution $v_{NN}$ here. Such discrepancy terms are common in the computer model experiment literature \citep[e.g.,][]{gramacy2020surrogates}. In the context of B-PINNs, a discrepancy term can be added in the data model, e.g., $z_{t^i}(\bm{s}^i) = v_{NN,t^i}(\bm{s}^i;\bm{\theta}_W) + \delta_{t^i}(\bm{s}^i;\bm{\theta}_\delta) + \epsilon_{t^i}(\bm{s}^i)$, where $\delta_{t^i}(\bm{s}^i;\bm{\theta}_\delta)$ is a discrepancy term that depends on parameters $\bm{\theta}_\delta$. In the statistical surrogate and mechanistically-motivated statistics literature, this discrepancy is typically modeled as a GP, a functional (basis) expansion, or a machine learning surrogate. In the B-PINN context, it is natural to model the discrepancy as an additional (separate) neural network \citep[e.g.,][]{zou2024correcting}, with the discrepancy used to correct for the misspecification of the PDE when observing PDE residuals. In either case, one must be careful to ensure that there is sufficient prior information regarding the spatial (and possibly temporal) scale of the discrepancy process to ensure identifiability of the solution (see our example in Section \ref{sec:simul} below for more discussion).

Although the B-PINN framework is an increasingly useful paradigm to account for uncertainty in PINN applications, there are still potential limitations in its lack of an explicit hierarchical formulation, lack of consideration of covariate effects, lack of accommodation of dependent random effects, focus on normal data distributions, and lack of more complicated measurement/sampling errors.  In the next section, we show how the B-PINN fits into a more general deep BHM framework for dynamic spatio-temporal processes that can accommodate these concerns.

\section{Bayesian Hierarchical Model PINNs (BHM-PINNs)}
\label{sec:BHM-PINN}

A strength of an overview is drawing generalizations and connections to a larger body of work.
As part of this overview of PINNs, we show here that the B-PINN \citep{yang2021b} fits naturally into the BHM framework that has long been used to model dynamic spatio-temporal processes in statistics \citep[e.g., see][for a review]{cressie2011statistics}. It is important to note that this is not a new methodology, but simply a new use case of the spatio-temporal BHM framework. We denote this a {\it BHM-PINN} and show it is a flexible approach to bring mechanistic-inspired PDE dynamics to spatio-temporal modeling in a paradigm that can account for non-normal data distributions, covariate effects, and residual spatial or spatio-temporal processes. This is analogous to how mechanistic-informed models are included in physical-statistical BHMs. Thus, the BHM-PINN described in the following subsections generalizes both the traditional BHM spatio-temporal physical-statistical modeling framework and the PINN modeling framework.

\subsection{Data Models}
Assume we have up to $p$ different datasets at time $t$ given by $\{\bm{z}_{1t},\ldots,\bm{z}_{pt}\}$ that are all thought to be informative about our latent process of interest, $\bm{u}_t$. As is common in spatio-temporal statistics, we assume these datasets are conditionally independent given the latent spatio-temporal process. Specifically,
$
\bm{z}_{jt}| \bm{u}_t, \bm{\theta}_{zj} \sim  \;\; \mathcal{D}(h_j(\bm{u}_t); \bm{\theta}_{zj}),$ 
where for $j=1,\ldots,p$, $\mathcal{D}(h_j(\bm{u}_t);\bm{\theta}_{zj})$ is some data distribution (e.g., normal, Poisson, log-normal, etc.) that is conditioned on the latent spatio-temporal process, $\bm{u}_t$, and parameters $\bm{\theta}_{zj}$. Here, $h_j( \;)$ is a mapping function between the latent process and the data that can account for missing observations, nonlinear transformations, and change-of-support (spatial resolution) \citep[for more background see][]{cressie2011statistics}. For instance, in the example of Section \ref{sec:simul}, the function $h(.)$ implicitly is a composition of an incidence matrix (to account for missing data locations) and exponentiation (to provide a link function for the Poisson distribution, which requires a positive rate parameter); in other words, this mapping can accommodate a link function in a generalized linear model context. As discussed in the context of physical-statistical models in Section \ref{sec:MechMotMods}, data models can accommodate multiple observational data sources as well as mechanistic model output. In addition, as discussed in the context of B-PINNs in Section \ref{sec:B-PINN}, these data models can accommodate ``pseudo-data'' for PDE residuals, boundary conditions, and initial conditions.

\subsection{Process Models}
We let the latent process part of the model be decomposed into three terms; an overall time-varying spatial mean, a PINN dynamic process, and a residual (discrepancy) process, respectively as
\begingroup
\setlength{\abovedisplayskip}{3pt} 
\setlength{\belowdisplayskip}{3pt} 
\begin{equation}
\bm{u}_t = \bm{\mu}_t + {\bm{v}}_{NN,t} + \bm{\delta}_t,
\label{eq:genprocess}
\end{equation}
\endgroup
where we are interested in modeling the latent process at a fixed set of $n$ locations given by the $n \times 1$ vector ${\bm u}_t$. Then, $\bm{\mu}_t$ is the $n$-dimensional time-varying spatial mean, $\bm{v}_{NN,t}$ is the PINN at the $n$ locations of interest at time $t$, and $\bm{\delta}_t$ is the discrepancy term. Each of these terms can be specified in multiple ways as described below.

\subsubsection{Mean} 
The process mean $\bm{\mu}_t$ can accommodate covariates or other ``fixed effects''; e.g., $\bm{\mu}_t = \bm{X}_t \bm{\beta}_t$, where $\bm{X}_t$ is an $n \times q$ covariate matrix with potentially time-varying covariates, and $\bm{\beta}_t$ is a $q$-dimensional vector (that may be time varying).  In this case, we would specify a prior distribution $[\bm{\beta}_t]$ (which depends on the application; e.g., see the simulation example in the next section).  We use the term ``covariates'' quite loosely here, and note that $\bm{X}_t$ can accommodate seasonality, temporal trends, one hot encoding, exogenous covariate effects, etc.  In some cases, one may not need this term in the model, or it might not depend on time (e.g., it could represent a simple constant offset).

\subsubsection{PINN Dynamic Process}\label{sec:latentDynProc}
The dynamic component of the model is the most important as it represents the underlying spatio-temporal mechanism driving the process. This is the main place where we impart mechanistic information via the PINN as described in Section \ref{sec:pinn} or B-PINN as in Section \ref{sec:B-PINN}, denoted as ${\bm{v}}_{NN,t}$. Note that we are not explicitly writing out the initial and boundary conditions, but those can be incorporated in the PINN as described previously. Recall, the PDE constraint can be incorporated into ``observations'' of the PDE residuals (\ref{eq:PDEresidual}) as before (e.g., we can include a data model for these residuals and specify the observations to be zero unless there is a known PDE forcing term) or we can specify a prior on the time derivative of the PDE. This then accommodates the PDE dynamical structure and we note there are additional $\bm{\theta}_m$ parameters that come in the residual process from the PDE. As with the B-PINN framework, the residual variance term ($\sigma^2_r$) allows one to control the degree to which the neural network is constrained to follow the underlying PDE. 

\subsubsection{Discrepancy Process}
The discrepancy process term $\bm{\delta}_t$ is important because it respects the fact that our latent dynamical model represented by the process ${\bm{v}}_{NN,t}$ surely does not completely account for all of the latent spatio-temporal dependence in real-world data (i.e., ``all models are wrong''). This term could be accommodated by another NN, say $\bm{\delta}_t = \bm{\delta}_{NN,t}(\cal{S};\bm{\theta}_{\delta})$ as in \citet[][]{zou2024correcting}.  Alternatively, it could be a time-varying stochastic process, e.g., $\bm{\delta}_t = f_{\delta}(\bm{\delta}_{t-1};\bm{\theta}_{\delta})$ such as the residual autoregresssive process to accommodate small scale dynamics in \citet{wikle2001spatiotemporal}, or simply a time-independent spatial GP. More generally, this term could be represented by any spatio-temporal random effect distribution, say $\bm{\delta}_t \sim [\bm{\theta}_{\delta}]$.

\subsection{Parameter Models}
Depending on the models selected in the data and process stages, one can have a variety of parameters that require prior distributions, e.g.,
$[\bm{\beta}_t | \bm{\theta}_{\beta}][\bm{\theta}_w][\bm{\theta}_{\delta}]$, etc.
Typically, we require regularization for these parameters as there are likely to be quite a lot of them, or as mentioned in the context of physical-statistical models, we might have mechanistic information for processes (e.g., parameters must obey a turbulent scaling relationship as in \citet[][]{wikle2001spatiotemporal}). In general, many of the prior specifications that have been used for mechanistically motivated BHM models (for some examples, see \cite{wikle2003hierarchical, pagendam2014assimilating, zammit2014resolving}) carry over to BHM-PINNs because the BHM-PINN takes the same form at the parameter level. In the B-PINN context, \cite{HBPINN} consider a hierarchical structure on parameters for a gear-grinding physics model in which correlation is allowed between parameters (we note, however, that this hierarchical framework is solely for the parameters of the physics model, distinct from the BHM-PINN approach). 

Deep BHMs as described here have multiple hyperparameters. These correspond to specifications of the parameters for the priors in the previous stages as well as model architecture (e.g., number of hidden units, levels, etc.) in the case of B-PINNs. Depending on the application, these hyperparameters can be given vague proper prior distributions or are fixed (i.e., point mass priors) and the sensitivity to the choice is examined. 

\subsection{Implementation Considerations}\label{sec:computation}

As with many mathematical and statistical models, model identifiability must be be addressed when implementing a BHM-PINN, specifically with regard to the NN and discrepancy terms.
For example, in (\ref{eq:genprocess}), if $\bm{\delta}_t$ is parameterized using a GP, there is potential for confounding between this term and the PINN term, ${\bm v}_{NN,t}$. 
One way to mitigate this issue is to restrict the spatial (temporal, or spatio-temporal) range of the GP to operate on a different scale than the dynamics imposed on $\bm{v}_{NN,t}$.
This can be accomplished by an informative prior distribution on the argument(s) of the kernel function for the GP.

All the implementation challenges for B-PINNs are present for BHM-PINNs. 
For example, it is not clear how to select the NN architecture in general, but see  \cite{toscano2025pinns} for a review of various NN architectures that have been considered in the context of PINNs.  Simple NNs, such as narrow 3-layer feed-forward NNs, are often used to fit B-PINNs.
Not only is this more computationally tractable than a more complex model, it is also related to the bias-variance trade-off, where wide and deep NNs can be over-parameterized and have the potential to overfit. 
Empirical results often suggest narrow and shallow feed-forward NNs result in the best performance, although this may not be the case for complex dynamics. 

The primary computational tool that has been used for posterior inference in Bayesian hierarchical models is MCMC, such as Gibbs sampling (and variants like Metropolis-within-Gibbs sampling) \citep{brooks2011handbook}. Modern gradient-based methods, such as Hamiltonian Monte Carlo \citep[HMC;][]{neal2011mcmc} and extensions like the No-U-Turn Sampler \citep{10.5555/2627435.2638586}, have become prevalent in general for Bayesian inference. In the B-PINN literature \citep{yang2021b}, the two primary computational tools are HMC and variational inference \citep{blei2017variational}. In the case of variational Bayes, the mean-field approximation (i.e., product of independent normals) for posterior neural network weights might not be realistic in the sense that correlations (which are likely to exist in neural network parameters) are neglected, although variational GPs \citep{tran2015variational} can help alleviate this issue. \citet{yang2021b} found gradient-based MCMC methods often result in the best performance for B-PINNs and for our example presented below in Section \ref{sec:simul}, we use this approach.

In the context of the discrepancy term, we note that 
GP models can be computationally difficult due to cubic complexity of inversion and determinant calculations. This is a well-researched topic in spatial statistics and there are several ways to mitigate this issue  \citep[e.g., see the review in][]{heaton2019case}.


\section{Simulation Experiment: BHM-PINN}
\label{sec:simul}
We consider a simulation experiment to illustrate a specific BHM-PINN.
In particular, we consider a latent mechanistic dynamic process described by a one-dimensional (in space) Burgers' equation \citep{bateman1915some,burgers1948mathematical}, which is a classic nonlinear PDE given by
\begin{equation}
\frac{\partial {v}(s,t)}{\partial t} 
+ {v}(s,t) \frac{\partial {v}(s,t)}{\partial s} 
- \lambda \frac{\partial^2 {v}(s,t)}{\partial s^2} = 0,
\label{eqn:burgers1D}
\end{equation}
where $\lambda$ is the diffusion coefficient. The Burgers' equation is often used to illustrate dynamics because it is a very simple nonlinear advection-diffusion representation of a fluid dynamics problem that can model features like nonlinear wave propagation and shock formation while remaining relatively easy to analyze and solve. Here, we also include a fixed-effects mean term in the process representation, along with a Gaussian process discrepancy to reflect a situation where the true process is not just a known mechanistic model. We then assume this process corresponds to the latent mean of a Poisson count data generating model. The details of the simulation are given below.

\begin{figure}[t]
\centering
\includegraphics[width=1\linewidth]{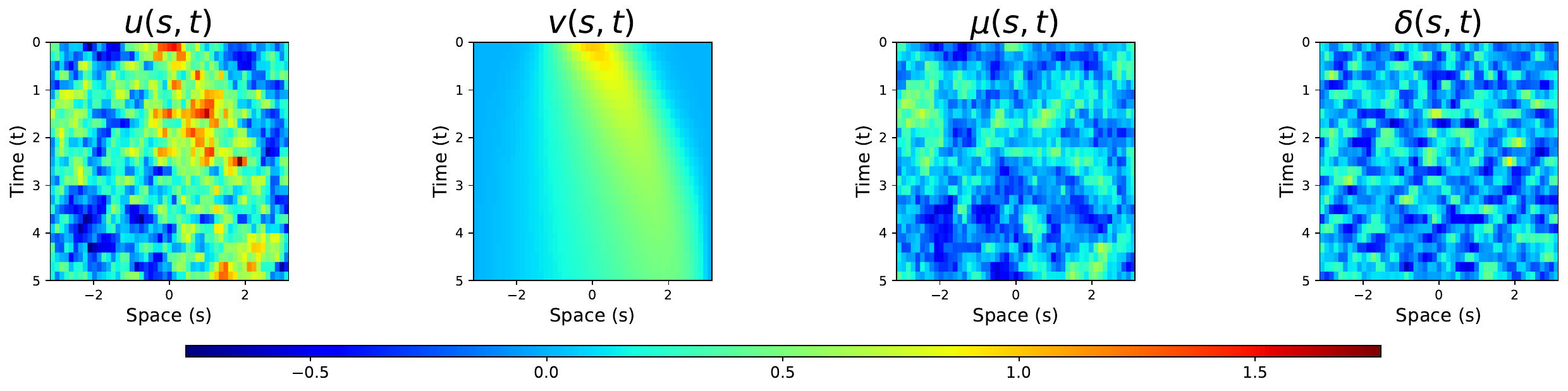}
  \caption{Simulated latent process $u(s,t)$ (left) and components over one-dimensional space and time: Burgers' equation ${v}(s,t)$ (second from left), mean process $\mu(s,t)$ (second from right), and discrepancy process $\delta(s,t)$ (right).}
    \label{data_breakdown}
\end{figure}

\begin{figure}[t]
\centering
\includegraphics[width=0.6\linewidth]{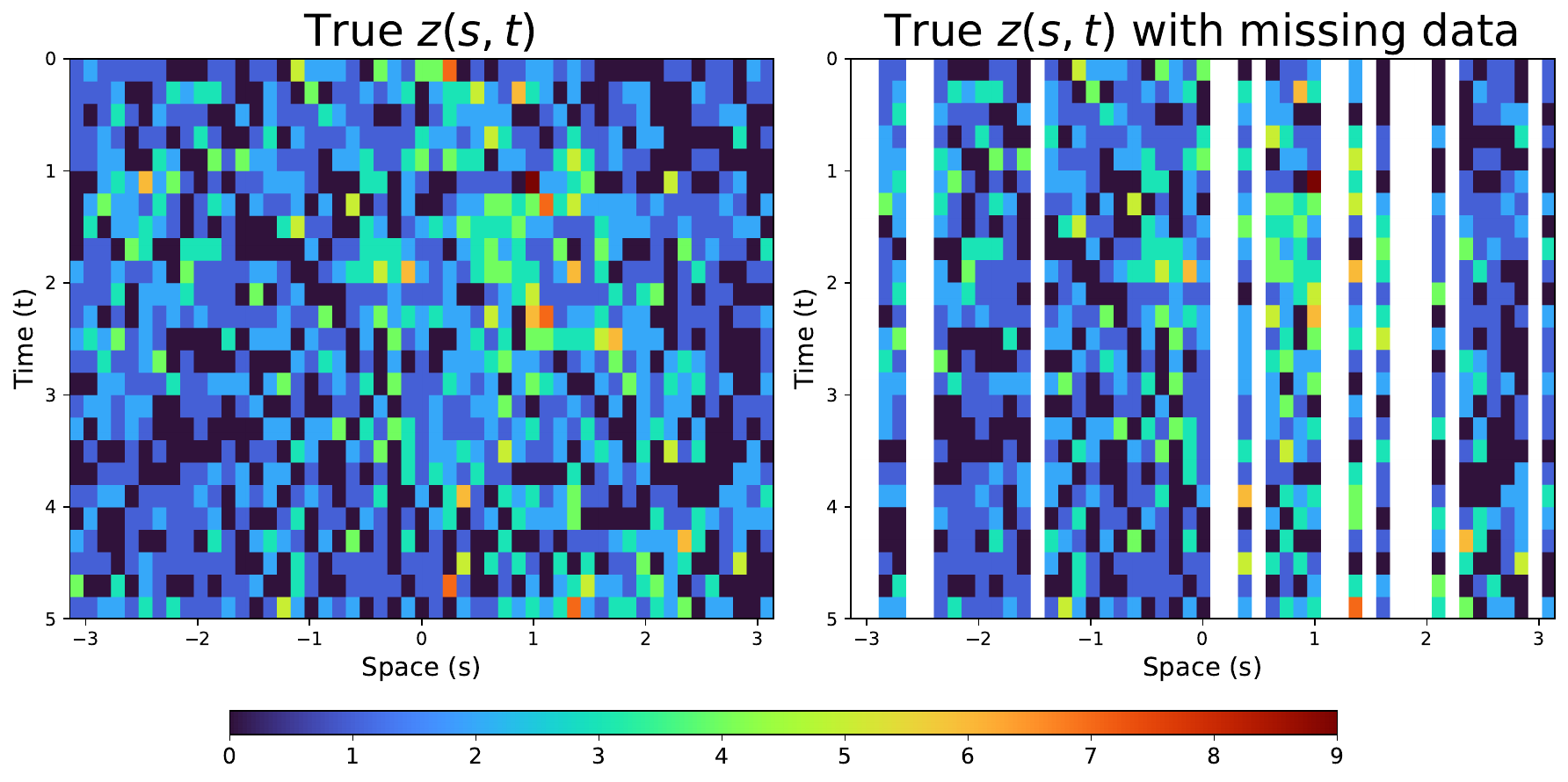}
  \caption{Left: Simulated $\bm{z}_t$, Right: Missing data in $\bm{z}_t$ highlighted in white.}
    \label{data_plot}
\end{figure}

\begin{enumerate}
\item {\bf Domain:} We consider a one-dimensional equally spaced spatial grid with $n=51$ locations in the domain $s \in [-\pi, \pi]$ and $T=25$ equally distributed time points, $t \in [0,5]$.

\item {\bf Process Mean ($\mu(s,t)$):} We consider two covariates in the mean process, $\bm{x}(s,t) =(x_1(s,t),x_2(s,t))^{\top}$, which are simulated from a zero-mean spatio-temporal GP with exponential covariance function $k_x((s,t),(s',t'))=0.1\cdot \exp\bigg(-\frac{\sqrt{(s - s')^2 + (t - t')^2}}{ \ell_s} \bigg)$ where $\ell_s$ is set to $1/20$ the maximum observed spatio-temporal distance across all location-time pairs. Specifically, we choose $\ell_s = 8.03/20$ and let $\bm{\beta}=(0.6,-0.4)^{\top}$. This is shown in the third panel of Figure \ref{data_breakdown}.

\item {\bf Latent Dynamic Process ($v(s,t)$):} For each spatial and temporal location, we generated the PDE solution to Burgers' equation (\ref{eqn:burgers1D}) numerically (with a spectral solver), denoted as ${v}(s,t)$. This numerical solution used initial condition ${v}(s,0) = \exp(-s^2)$ for $s \in (-\pi, \pi)
$ and 0 otherwise, with boundary conditions ${v}(-\pi,t) = {v}(\pi,t) = 0$. This is shown in the second panel of Figure \ref{data_breakdown}.

\item {\bf Discrepancy Process ($\delta(s,t)$):} For each time point a spatial discrepancy field was drawn as an independent, zero-mean Gaussian process, i.e., $\delta(s,t)\sim \text{GP}({0},k_\delta(s,s'))$, where  $k_\delta(s,s')=\sigma^2_{\delta} \exp\bigg(-\frac{(s-s')^2}{2 \ell_{\delta}^2} \bigg)$ is a squared-exponential covariance function. Specifically, we select $\sigma^2_{\delta}=0.05$, and $\ell_{\delta}=0.15$. This is shown in the fourth panel of Figure \ref{data_breakdown}.

\item {\bf Simulated Process ($u(s,t)$):} The simulated process at each spatial and temporal coordinate is then given as the sum of the mean, latent dynamic process, and discrepancy: $u(s,t) := \bm{x}(s,t)^{\top} \bm{\beta} + {v}(s,t) + \delta(s,t)$.  This is shown in the fourth panel of Figure \ref{data_breakdown}.

\item {\bf Simulated Observations ($z(s,t)$):} We then simulated observations from a random Poisson distribution, $z(s,t) \sim \mbox{Poisson}(\textrm{rate} = \exp(u(s,t)))$ (Figure \ref{data_plot} left panel).
We assumed that observations at 30\% of the space coordinates are missing at random for all times (Figure \ref{data_plot} right panel).
Note, data missing over space for all times is a realistic scenario for simulating environmental processes with incomplete observation networks and can be more challenging to model than cases where data are missing completely at random in space and time.
\end{enumerate}

We consider the following BHM to fit these simulated data. First, the data, process, and discrepancy models are given by:
\begingroup
\setlength{\abovedisplayskip}{3pt} 
\setlength{\belowdisplayskip}{3pt} 
\begin{align*}
\text{Data Model}: & \quad \bm{z}_t \sim \mbox{Poisson}(\exp(\bm{H}_t \bm{u}_t)) \\
\text{True Process}:& \quad u(s,t)= \bm{x}(s,t)^{\top} \bm{\beta} + {v}_{NN}(s,t) + \delta(s,t) \\
\text{Latent dynamics ($v_{NN}(s,t;\bm{\theta}_W)$}:
& \quad \frac{\partial {v}_{NN}(s,t)}{\partial t} \sim \text{N}(v_{NN}(s,t) \frac{\partial {v}_{NN}(s,t)}{\partial s} - \lambda \frac{\partial^2 {v}_{NN}(s,t)}{\partial s^2}, \sigma^2_{r})\\
& \quad {v}_{NN}(s,0) \sim \text{N}(\exp(-s^2), \sigma^2_{ic}) \quad \text{(Initial condition)} \\
& \quad {v}_{NN}(-\pi,t), {v}_{NN}(\pi,t) \sim \text{N}(0, \sigma^2_{bc}) \quad \text{(Boundary condition)}\\
\text{Discrepancy process}:& \quad \delta(s,t)\sim \text{GP}(0,k_\delta(s,s')) \\
&\quad k_\delta(s,s')=\sigma^2_{\delta} \exp\bigg(-\frac{(s-s')^2}{2 \ell_{\delta}^2} \bigg).
\end{align*}
In this case, $\bm{H}_t$ is an incidence matrix that accommodates missing observations \citep[e.g., see the discussion in][]{wikle2019}. In addition, the following prior distributions are specified:
\begingroup
\setlength{\abovedisplayskip}{3pt} 
\setlength{\belowdisplayskip}{3pt} 
\begin{alignat}{2}
\bm{\beta} &\sim \text{N}(\bm{\mu}_{\bm{\beta}},c_{\bm{\beta}} \bm{I} ), \qquad && \lambda \sim \text{N}(\mu_{\lambda},\sigma^2_{\lambda})\cdot \text{I}( 0 \leq\lambda < \infty ) \nonumber\\
\ell_{\delta} &\sim \text{N}(\mu_{\ell},\sigma^2_{\ell})\cdot \text{I}( \ell_{\ell} \leq\ \ell_{\nu} < u_{\ell} ), \qquad
&& \sigma^2_{\delta} \sim \text{N}(\mu_{\delta},\gamma
^2_{\delta}) \cdot \text{I}(\ell_{\delta}<\sigma^2_{\delta}<u_{\delta})  \nonumber \\ 
\bm{\theta}_W  & \sim \text{N}(\bm{\mu}_{W},c_{W} \bm{I}) \nonumber ,
\end{alignat}
\endgroup
where relatively informative priors are assigned for $\sigma^2_{\delta}$ and $\ell_{\delta}$ with $\mu_{\delta}=\mu_{\ell}=0,\gamma^2_{\delta}=\sigma^2_{\ell}=1$, and truncation bounds $\ell_{\delta}=0.02, u_{\delta}=0.1, \ell_{\ell}=0.05,$ and $u_{\ell}=0.2$. These informative priors on the discrepancy parameters address the potential identifiability issue between the NN and discrepancy as well as the ability to recover Gaussian process covariance parameters in a fixed domain setting (an issue which has been frequently noted in the spatial statistics literature, e.g.,  \cite{gelfand2010handbook}). In contrast, we assign relatively non-informative priors to $\bm{\beta}$, $\lambda$, and the neural network weight and bias parameters $\bm{\theta}_{W}$, with $\bm{\mu}_{\bm{\beta}}=\bm{\mu}_W=\bm{0}$, $c_W=\sigma^2_{\lambda}=1$, $\mu_{\lambda}=0$, and $c_{\bm{\beta}}=10$. The neural network for ${\bm{v}}_{NN,t}$ consists of $L=3$ hidden layers, each with 16 hidden units.

As discussed in Section \ref{sec:B-PINN}, the specification of the B-PINN model in this BHM uses ``pseudo-data'' for the PDE residuals, boundary conditions, and initial conditions. Specifically, we can write the following conditional data distributions that have the effect of updating the prior distributions for weights $\bm{\theta}_W$.
\begin{align}
[D_r|\sigma^2_r,\bm{\theta}_W,\lambda]&=\text{N}(0;[\frac{\partial v_{NN}} {\partial t}+ v_{NN}\frac{\partial v_{NN}}{\partial s} - \lambda \frac{\partial ^2 v_{NN}}{{\partial s^2}}](s, t; \bm{\theta}_W),\sigma^2_r), \quad (\text{Burgers' equation})  \nonumber\\
[D_{bc+}|\sigma^2_{bc},\bm{\theta}_W]&=\text{N}(0;v_{NN}(\pi,t; \bm{\theta}_W),\sigma^2_{bc}),\quad (\text{Right boundary condition})  \nonumber \\
[D_{bc-}|\sigma^2_{bc},\bm{\theta}_W]&=\text{N}(0;v_{NN}(-\pi,t; \bm{\theta}_W),\sigma^2_{bc}),\quad (\text{Left boundary condition}) \nonumber \\
[D_{ic}|\sigma^2_{ic},\bm{\theta}_W]&=\text{N}(\exp(-s^2); v_{NN}(s,0; \bm{\theta}_W), \sigma^2_{ic}), \quad (\text{Initial condition}) \nonumber, 
\end{align}
where $\sigma^2_r$, $\sigma^2_{bc}$, and $\sigma^2_{ic}$ govern the extent to which the neural network $v_{NN}$ is constrained to satisfy Burgers' equation along with the initial conditions and boundary conditions. Assuming reasonably strong prior information for the PDE, boundary conditions, and initial conditions, we set $\sigma^2_r = 0.05^2$ and $\sigma^2_{bc} = \sigma^2_{ic} = 0.01^2$. We re-emphasize that assuming the residual data are zero (e.g., $D_r = 0$) further constrains the prior on $\bm{\theta}_W$ and these residuals are not actual measurements of the process.

\begin{figure}[t]
\centering
\includegraphics[width=1\linewidth]{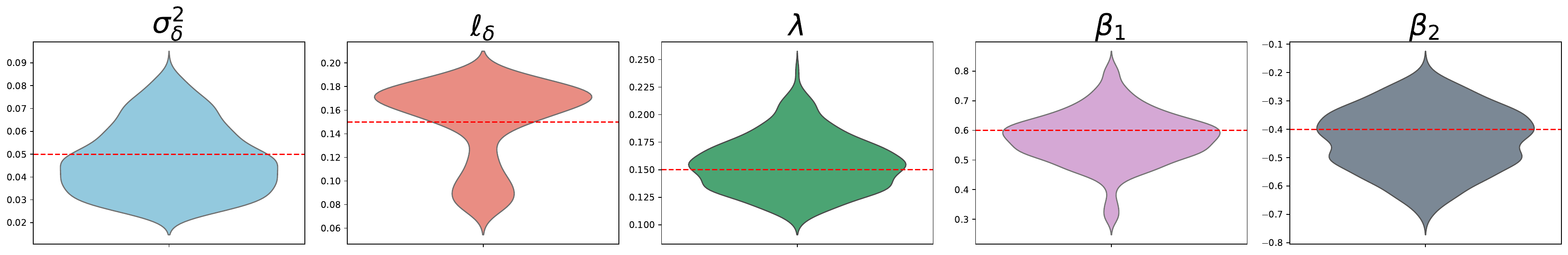}
  \caption{Violin plots of the BHM model parameter posterior distributions, with the true value indicated by a dashed red line. From left to right, the plots correspond to $\sigma^2_\delta$, $\ell_\delta$, $\lambda$, $\sigma^2_z$, $\beta_1$, and $\beta_2$. }
    \label{violin_plot}
\end{figure}

After specifying the prior distributions and the neural network architecture, we draw 2,000 posterior samples using HMC with the No-U-Turn Sampler \citep{10.5555/2627435.2638586}, discarding the first 1,000 as burn-in and using a tree depth of 10. This is implemented using the ``NumPyro" \citep{phan2019composable} and ``JAX" \citep{jax2018github} packages in Python and takes approximately 45 minutes on an Apple MacBook M4 Pro with 128GB RAM.

\subsection{Results}
We first evaluate whether the posterior samples generated from the No-U-Turn sampler produce accurate estimates of the true simulation parameters. Figure \ref{violin_plot}  presents violin plots for the parameters, excluding the neural network weights and biases $\bm{\theta}_W$, which are not the primary focus. Although some parameters show slight estimation bias, their credible intervals include the true values and are indicative of Bayesian learning.

\begin{figure}[t]
\centering
\includegraphics[width=1\linewidth]{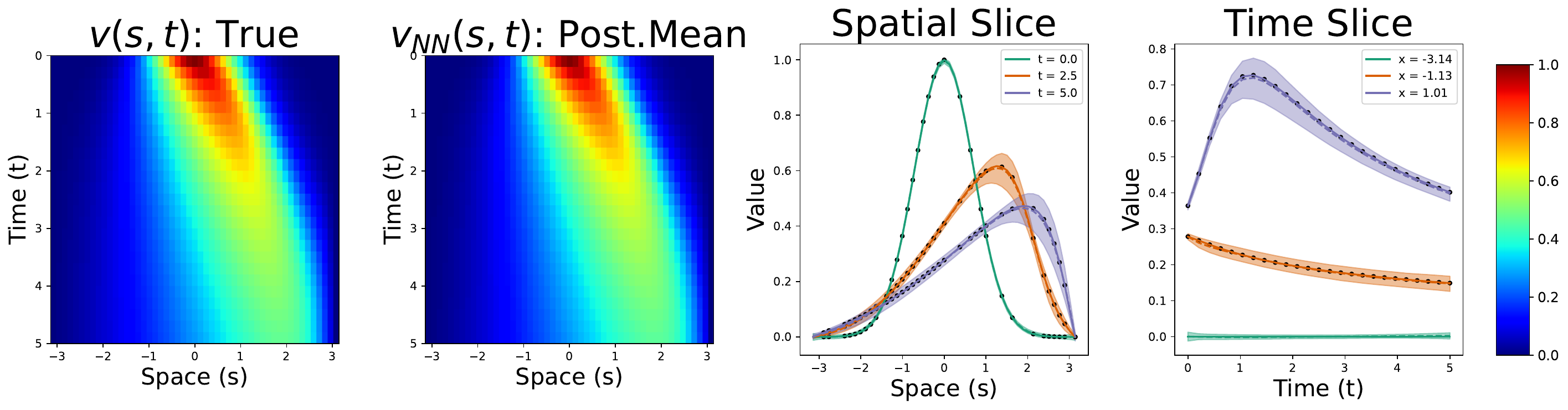}
    \caption{Left: true $\bm{v}_t$; Second from the left: predicted $v_{NN}(s,t)$; Second from the right: posterior means (dashed line) and corresponding 95\% credible intervals (shaded region) for predicted values of a marginal spatial slice of $\bm{v}_{NN,t}$ at times $t = 0$, $2.5$, and $5$ compared to the true process (solid line); Right: Posterior means (dashed line) and corresponding 95\% credible intervals (shaded region) for a marginal time slice of the predicted values of $\bm{v}_t$ at spatial locations $s = -3.14$, $-1.13$, and $1.01$ compared to the true process (solid line). In the right two plots, the dots correspond to spatial locations or time points when data are observed.}
    \label{burgers_plot}
\end{figure}

\begin{figure}[t]
\centering
\includegraphics[width=1\linewidth]{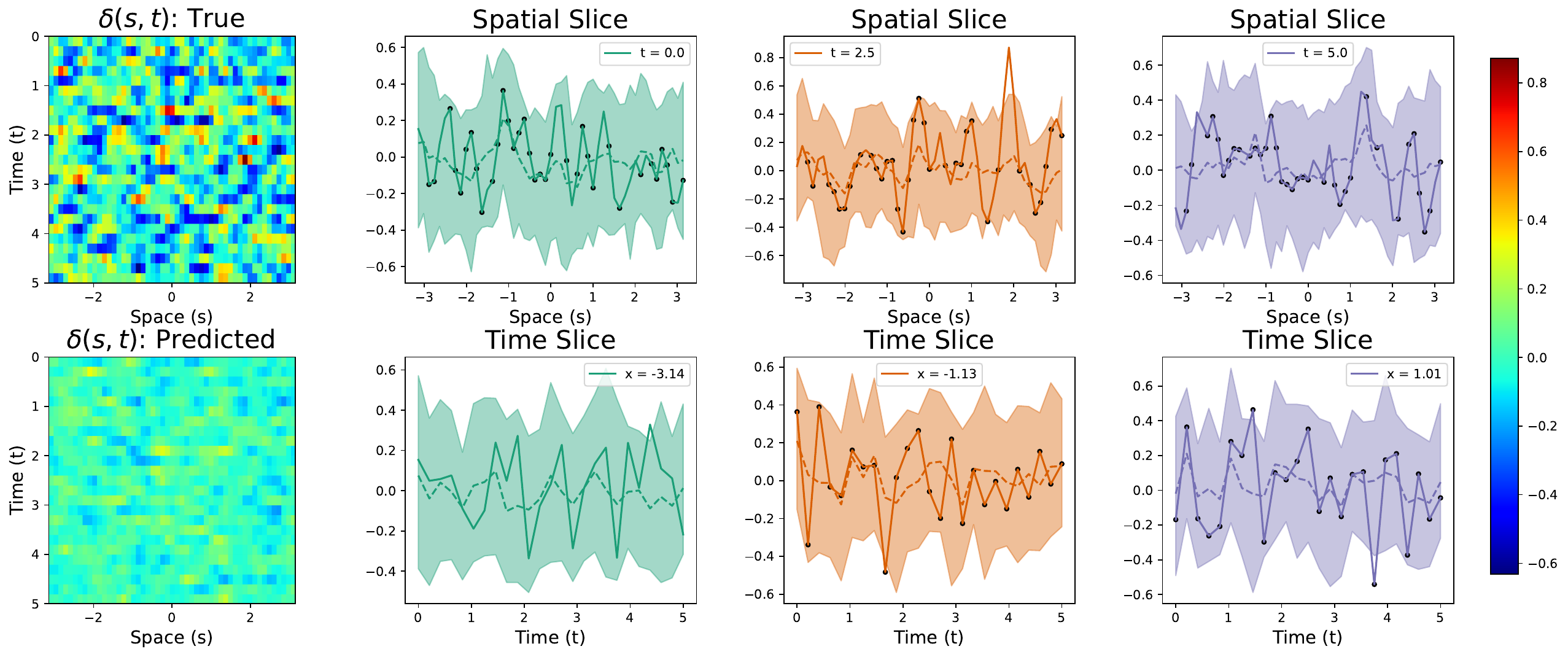}
    \caption{First Row, First Column: true values of the spatio-temporal discrepancy term $\bm{\delta}_t$; Second Row, First Column: corresponding posterior prediction of $\bm{\delta}_t$; First Row, Columns 2-4: spatial slice of posterior mean (dashed line) and 95\% credible intervals (shaded region) for times $t=0$, $2$, and $t=5$ compared to the true process (solid line); Second Row, Columns 2-4: time slice of posterior mean (dashed line) and 95\% credible intervals (shaded region) for spatial locations $s=-3$, $-1$, and 1 compared to the true process (solid line). For all plots in columns 2-4, the dots correspond to spatial locations or time points when data is observed.}
    \label{gp_plot}
\end{figure}
Figures \ref{burgers_plot} and \ref{gp_plot} compare the true values and predictions of ${\bm{v}}_{t}$ and $\bm{\delta}_t$, respectively. As shown in Figure \ref{burgers_plot}, the predicted ${\bm{v}}_{t}$ space-time field and spatial and temporal marginal plots align well with the true values, despite the substantial amount of missing data. This is one of the primary advantages of incorporating mechanistic information into the model. 
In contrast, Figure \ref{gp_plot} demonstrates that accurately predicting the discrepancy term, $\bm{\delta}_t$, is more challenging and subject to greater uncertainty, although the posterior mean (dashed line in the spatial and temporal slice plots) still reasonably matches the true process (solid line). Combining posterior samples from each term, Figure \ref{process_plot} shows the posterior mean of the overall process $\bm{u}_t$. While there are slight deviations from the true $\bm{u}_t$, the posterior mean successfully captures the underlying spatio-temporal pattern, demonstrating the flexibility of the model. 

We demonstrate a similar BHM-PINN for the case of normally-distributed data in the supplementary material. The results are quite comparable, if not slightly better, than those presented here for the Poisson data.
\begin{figure}[t]
\centering
\includegraphics[width=0.7\linewidth]{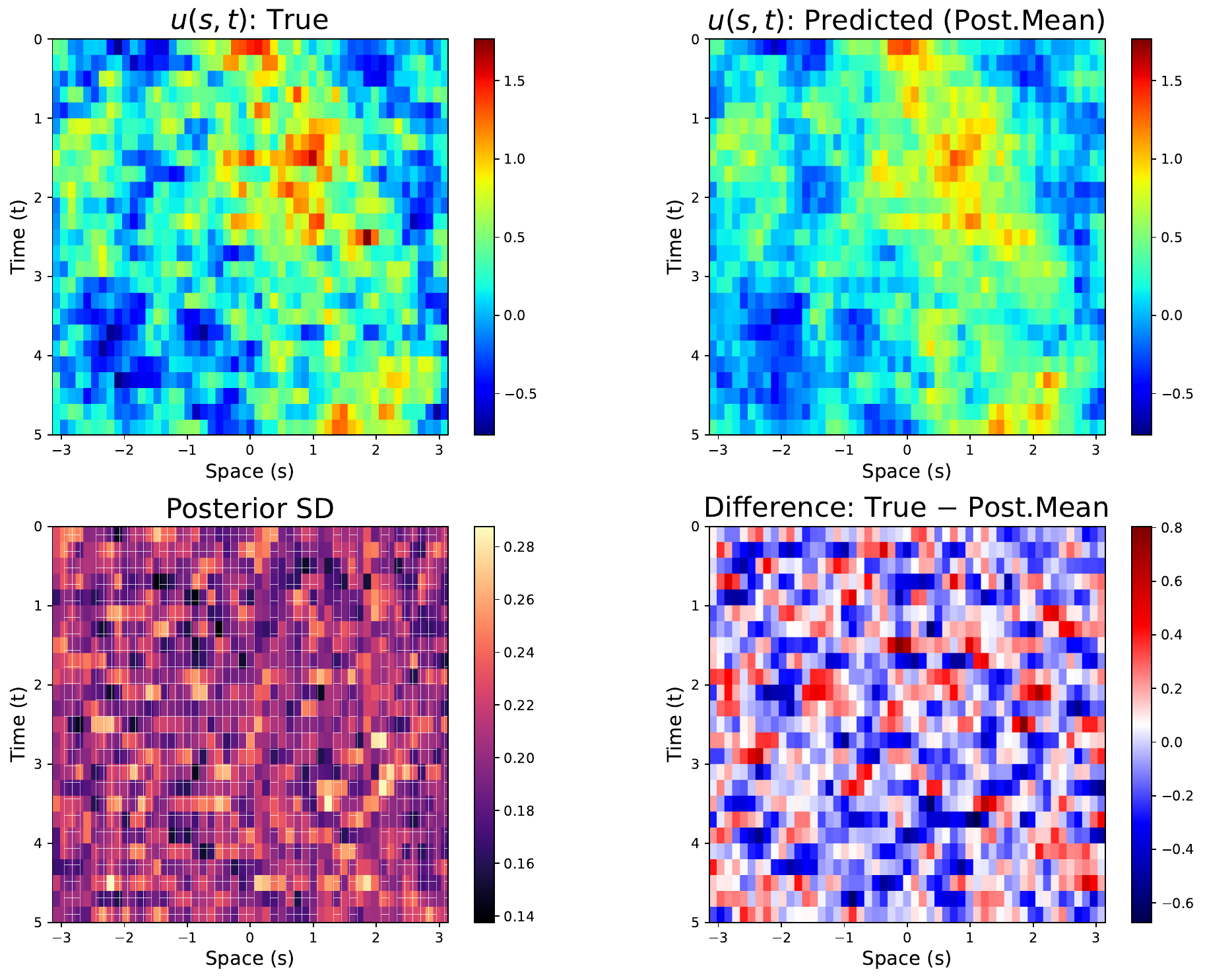}
  \caption{Top Left: true simulated $\exp(\bm{u}_t)$; Top Right: predicted posterior mean of $\exp(\bm{u}_t)$; Bottom Left: posterior standard deviation, with white rectangles indicating locations where data are observed; Bottom Right: difference between the true $\bm{u}_t$ and the posterior mean.}
    \label{process_plot}
\end{figure}
\section{Discussion}
\label{sec:Conclusion}
The incorporation of mechanistic information in data-driven models has been of interest to scientists for decades. There are numerous paradigms for such data-driven approaches, ranging from variational data assimilation, online filtering, physical-statistical models and probabilistic numerical methods.  Recently, there has been a resurgence of interest in these approaches because of the success of using deep neural networks with physical constraints \citep{Raissi2020, yang2021b, zou2024correcting}. Here, we review these PINN approaches and show some connections to the statistical approaches, followed by a generalization of B-PINNs to show that a general BHM framework can accommodate the B-PINN dynamical process but with the added flexibility to incorporate non-normal data models, covariate effects, and discrepancy processes. We illustrate such a BHM-PINN with the simulation of Poisson spatio-temporal data that are informed by the classic Burgers' nonlinear PDE, a covariate-based mean process, and a GP discrepancy term, and demonstrate we can recover important components of the model given uncertain and missing data. 

The main contribution of this work is to place the PINN and B-PINN in the context of more traditional statistical approaches to incorporate mechanistic information in spatio-temporal models. Importantly, as a generalization, we show that a B-PINN naturally fits into a BHM framework (i.e., a BHM-PINN), which can accommodate non-normal data, covariates, and random discrepancy terms. 

In some applications it maybe be preferable to embed a mechanistically-informed dynamical random spatio-temporal dynamical process as opposed to a neural network as in traditional physical-statistical BHMs. For example, \citet{wikle2001spatiotemporal} model 2-D spatio-temporal wind velocity components as a vector-autoregressive process that is motivated by an underlying set of (linear) PDEs, and  \citet{gopalan2019hierarchical} motivates the latent process by a numerical PDE solution summed with a stochastic discrepancy process. However, for general nonlinear PDEs there may not be a convenient autoregressive or general quadratic nonlinear \citep{wikle2010} representation of the latent dynamics. In such cases, a mechanistic-BHM using a B-PINN approach may be preferred since automatic differentiation can be used for general, nonlinear PDEs. This could also be an advantage of using a model such as the one in Section \ref{sec:BHM-PINN} over PNM methods that lack a neural network representation, in which case it is not obvious that automatic differentiation would apply.

Related to the previous point, in the classical DSTM BHM \citep[e.g.,][]{cressie2011statistics} it is often useful to consider the latent dynamics in a dimension reduced framework such as ${\bm v}_t = {\bm \Phi}{\bm \alpha}_t$ instead of neural network dynamics, ${\bm{v}}_{NN,t}$. In such a case, ${\bm v}_t$ is the $n$-dimensional latent dynamic process and ${\bm \Phi}$ is an $n \times n_\alpha$ dimensional basis function matrix with ${\bm \alpha}_t$ an $n_\alpha$-dimensional dynamic process where $n_\alpha < n$. For certain basis functions (e.g., Fourier) and systems, one can specify underlying mechanistic differential equations that govern the dynamics of ${\bm \alpha}_t$ \citep[e.g.,][]{cressie2011statistics}. This spectral approach could provide additional computational efficiency if the dynamics can be represented in a lower-dimensional manifold (reduced rank) basis expansion. 

From our exploration of these methods, in particular, the incorporation of a B-PINN within a more general BHM, there are several challenges with PINN-based models that suggest additional research. Of primary concern is the challenge of computational efficiency when scaling to a realistic spatio-temporal application. Although automatic differentiation is quite helpful for implementing PDE models, it can be slow, and there are well-known challenges to Bayesian learning of deep neural network parameters. In addition, real-world processes are often described by more complex dynamics than can be parameterized by a PDE (all models are wrong), and modeling the discrepancy becomes crucial. Our simulation studies have shown there can be significant identifiability issues with a flexible discrepancy process and the underlying mechanistic process without rough {\it a priori} guidance as to at least the space and time scales corresponding to each. It is also the case the data volume and orientation can make a difference in accurately estimating the various model components.  Lastly, as with all deep neural implementations, there is typically no {\it a priori} guidance to inform the model architecture.  These are very active areas of research and we are optimistic that these challenges can be met and these methods can be scaled to many realistic spatio-temporal problems.

\section*{Acknowledgments}

JSN was supported by the Director, Office of Science, Office of Biological and Environmental Research of the U.S. Department of Energy under Contract No. DE-AC02-05CH11231 and by the Regional and Global Model Analysis Program area within the Earth and Environmental Systems Modeling Program as part of the Calibrated and Systematic Characterization, Attribution, and Detection of Extremes (CASCADE) project. 

GG was supported by the Verification and Validation subprogram of the Advanced Simulation and Computing Program at the Los Alamos National Laboratory.

\newpage
\appendix
\section*{Supplementary Material}
\begin{figure}[t]
\centering
\includegraphics[width=1\linewidth]{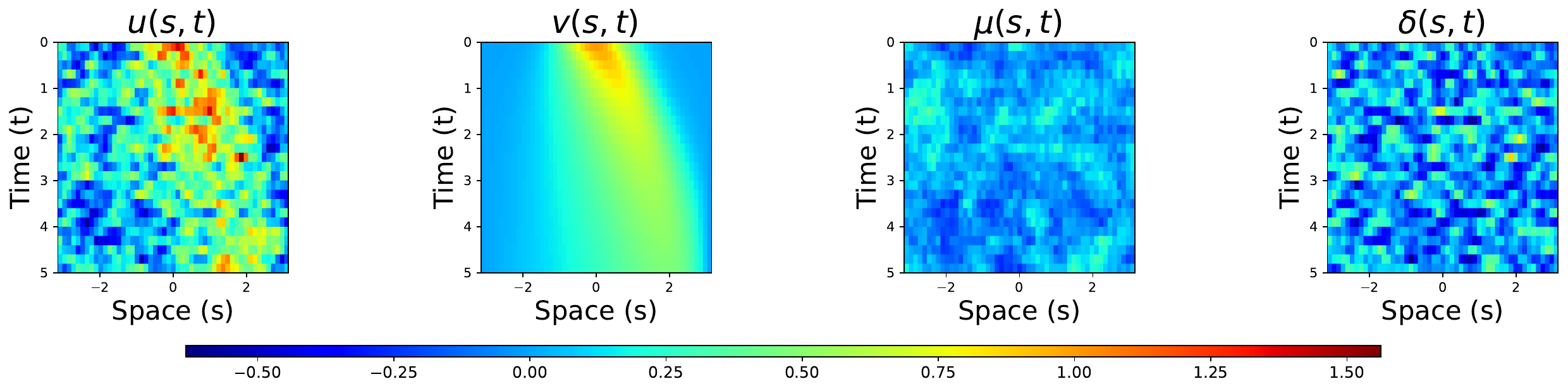}
  \caption{Simulated latent process $u(s,t)$ (left) and components over one-dimensional space and time: Burgers' equation ${v}(s,t)$ (second from left), mean process $\mu(s,t)$ (second from right), and discrepancy process $\delta(s,t)$ (right).}
    \label{supple_fig1}
\end{figure}
\begin{figure}[t]
\centering
\includegraphics[width=1\linewidth]{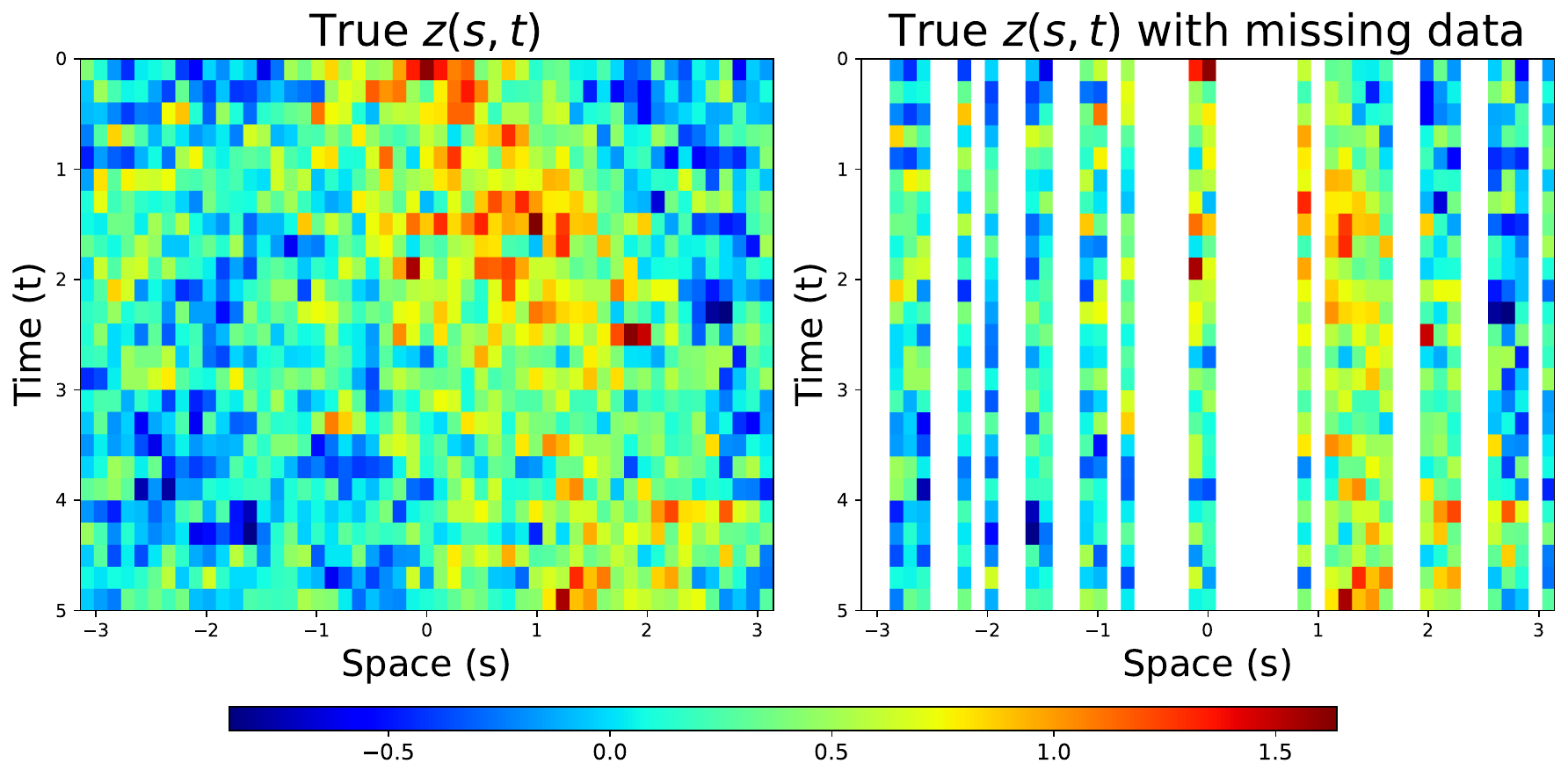}
  \caption{Left: Simulated $\bm{z}_t$, Right: Missing data in $\bm{z}_t$ highlighted in white.}
    \label{fig2_supple}
\end{figure}
We present a simulation experiment where the data model follows a Normal distribution as opposed to Poisson as in the main paper. Specifically, the details of the simulation are the same as those described in the manuscript, except for: 
\begin{enumerate}
\item {\bf Process Mean ($\mu(s,t)$):} We consider two covariates in the mean process, $\bm{x}(s,t) =(x_1(s,t),x_2(s,t))^{\top}$, which are simulated from a zero-mean spatio-temporal GP with exponential covariance function $k_x((s,t),(s',t'))=0.1\cdot \exp\bigg(-\frac{\sqrt{(s - s')^2 + (t - t')^2}}{ \ell_s} \bigg)$ where $\ell_s$ is set to $1/20$ the maximum observed spatio-temporal distance across all location-time pairs. Specifically, we choose $\ell_s = 8.03/20$ and let $\bm{\beta}=(0.3,-0.2)^{\top}$.
\item {\bf Simulated Observations ($z(s,t)$):} We then simulated observations from a Normal distribution, $z(s,t) \sim \text{N}( u(s,t),\sigma^2_d),$ where $\sigma^2_d=0.2$.
We assumed that 50\% of the space coordinates are missing.
\end{enumerate}

\begin{figure}[t]
\centering
\begin{subfigure}[b]{1\linewidth}
    \centering
    \includegraphics[width=1\linewidth]{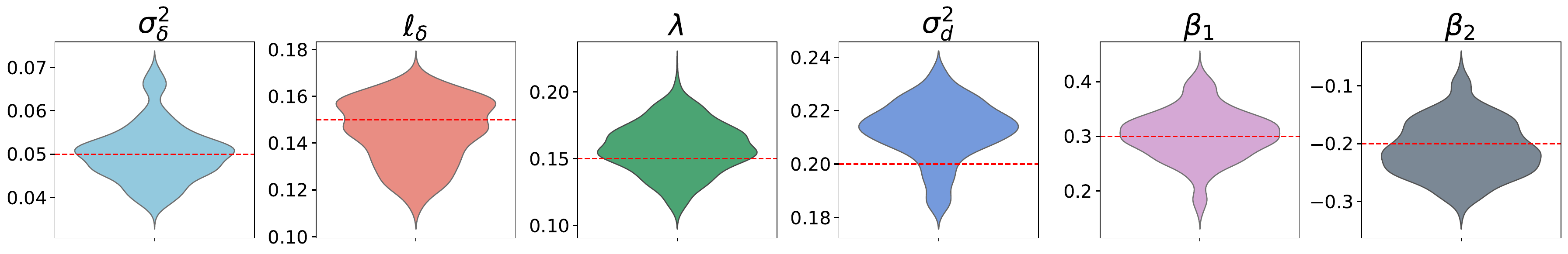}
    \caption{Violin plots of the BHM model parameter posterior distributions, with the true value indicated by a dashed red line. From left to right, the plots correspond to $\sigma^2_\delta$, $\ell_\delta$, $\lambda$, $\sigma^2_d$, $\beta_1$, and $\beta_2$. }
    \label{fig3_supple}
\end{subfigure}

\vspace{0.5cm}

\begin{subfigure}[b]{1\linewidth}
    \centering
    \includegraphics[width=1\linewidth]{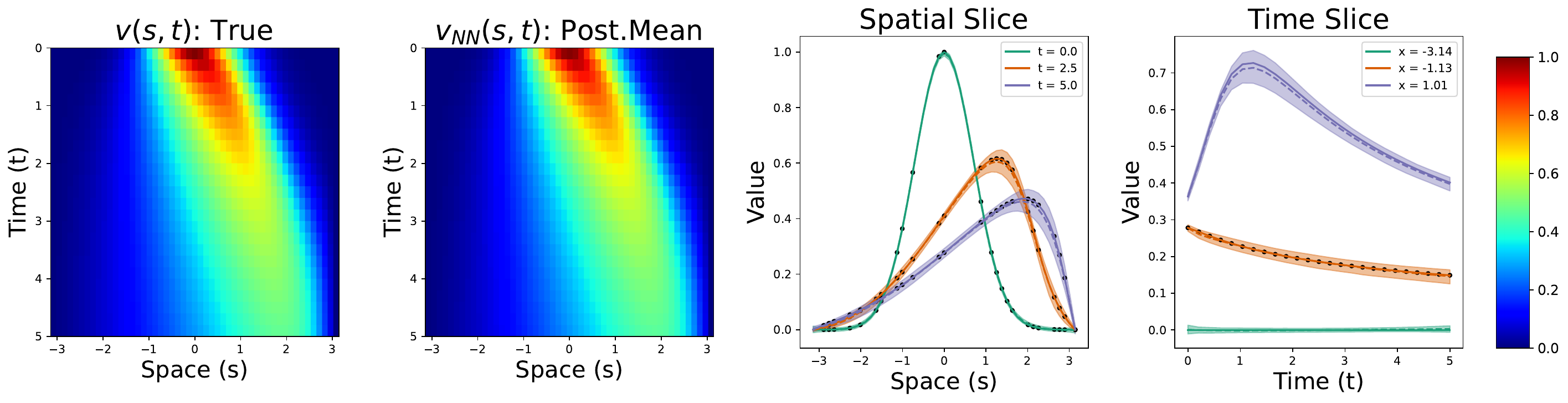}
    \caption{Left: true $\bm{v}(s,t)$; Second from the left: Posterior mean of $\bm{v}_{NN}(s,t)$; Second from the right: Posterior means (dashed line) and corresponding 95\% credible intervals (shaded region) for predicted values of a marginal spatial slice of $\bm{v}_{NN}(s,t)$ at times $t = 0$, $2.5$, and $5$ compared to the true process (solid line); Right: Posterior means (dashed line) and corresponding 95\% credible intervals (shaded region) for a marginal time slice of the predicted values of $\bm{v}_{NN}(s,t)$ at spatial locations $s = -3.14$, $-1.13$, and $1.01$ compared to the true process (solid line). In the right two plots, the dots correspond to spatial locations or time points when data are observed.}
    \label{fig4_supple}
\end{subfigure}
\caption{Results}
\label{fig_supple_combined}
\end{figure}
Figure \ref{supple_fig1} shows the simulated latent process and associated components and \ref{fig2_supple} shows the simulated data and the missingness used when fitting the model. Specifically, we consider the following BHM to fit these simulated data. First, using the notation defined in the main paper, the data and process models are given by
\begin{align*}
\text{Data Model}: & \quad \bm{z}_t = \bm{H}_t\bm{u}_t +\bm{\epsilon}_t,\quad \bm{\epsilon}_t\sim \text{N}(\bm{0},\sigma^2_d \bm{I}) \\
\text{True Process}:& \quad u(s,t)= \bm{x}(s,t)^{\top} \bm{\beta} + v_{NN}(s,t) + \delta(s,t),
\end{align*}
where $\bm{H}_t$ is an incidence matrix. 
The latent dynamic and discrepancy processes are specified as
\begin{align*}
\text{Latent dynamic process ($v_{NN}(s,t;\bm{\theta}_W)$}:
& \quad \frac{\partial {v}_{NN}(s,t)}{\partial t} \sim \text{N}(v_{NN}(s,t) \frac{\partial {v}_{NN}(s,t)}{\partial s} - \lambda \frac{\partial^2 {v}_{NN}(s,t)}{\partial s^2}, \sigma^2_{r})\\
& \quad {v}_{NN}(s,0) \sim \text{N}(\exp(-s^2), \sigma^2_{ic}) \quad \text{(Initial condition)} \\
& \quad {v}_{NN}(-\pi,t), {v}_{NN}(\pi,t) \sim \text{N}(0, \sigma^2_{BC}) \quad \text{(Boundary condition)}\\
\text{Discrepancy process}:& \quad \delta(s,t)\sim \text{GP}(0,k(s,s')) \\
&\quad k(s,s')=\sigma^2_{\delta} \exp\bigg(-\frac{(s-s')^2}{2 \ell_{\delta}^2} \bigg).
\end{align*}
The prior distributions are specified as
\begin{alignat}{2}
\sigma^2_d &\sim \text{Cauchy}(\mu_{d},\gamma_d) \cdot \text{I}(\ell_{d}<\sigma^2_d<u_d), \qquad
&& \bm{\beta} \sim \text{N}(\bm{\mu}_{\bm{\beta}},c_{\bm{\beta}} \bm{I} )  \nonumber\\
\lambda &\sim \text{N}(\mu_{\lambda},\sigma^2_{\lambda})\cdot \text{I}( 0 \leq\lambda < \infty ), \qquad
&& \sigma^2_{\delta} \sim \text{Cauchy}(\mu_{\delta},\gamma_{\delta}) \cdot \text{I}(\ell_{\delta}<\sigma^2_{\delta}<u_{\delta})  \nonumber \\ 
\ell_{\delta} &\sim \text{N}(\mu_{\ell},\sigma^2_{\ell})\cdot \text{I}( \ell_{\ell} \leq\ \ell_{\delta} < u_{\ell} ), \qquad
&& \bm{\theta}_W \sim \text{N}(\bm{\mu}_{W},c_{W} \bm{I}) \nonumber ,
\end{alignat}

\begin{figure}[t]
\centering
\includegraphics[width=1\linewidth]{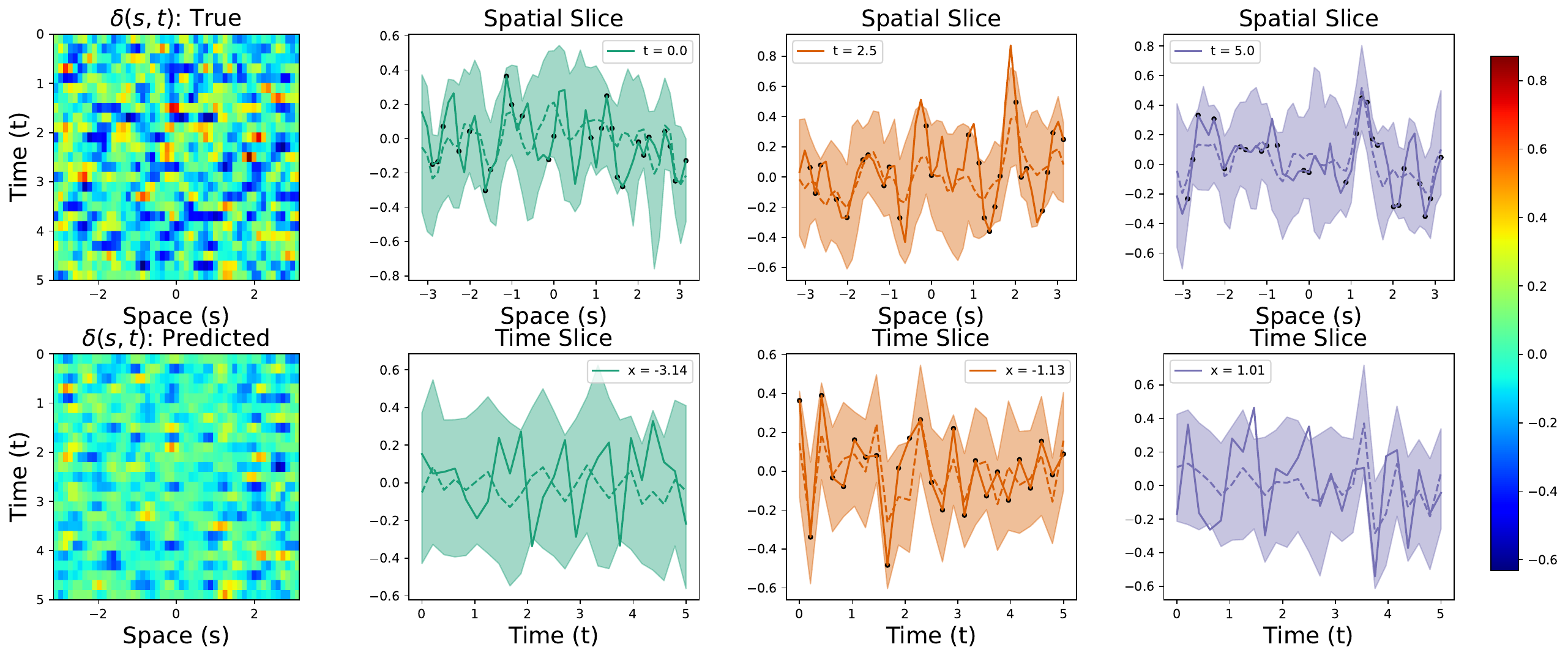}
    \caption{First Row, First Column: True values of the spatio-temporal discrepancy term $\bm{\delta}_t$; Second Row, First Column: Corresponding posterior prediction of $\bm{\delta}_t$; First Row, Columns 2-4: Spatial slice of posterior mean (dashed line) and 95\% credible intervals (shaded region) for times $t=0$, $2$, and $t=5$ compared to the true process (solid line); Second Row, Columns 2-4: Time slice of posterior mean (dashed line) and 95\% credible intervals (shaded region) for spatial locations $s=-3$, $-1$, and 1 compared to the true process (solid line). For all plots in columns 2-4, the dots correspond to spatial locations or time points when data is observed.}
    \label{fig5_supple}
\end{figure}

\begin{figure}[t]
\centering
\includegraphics[width=0.7\linewidth]{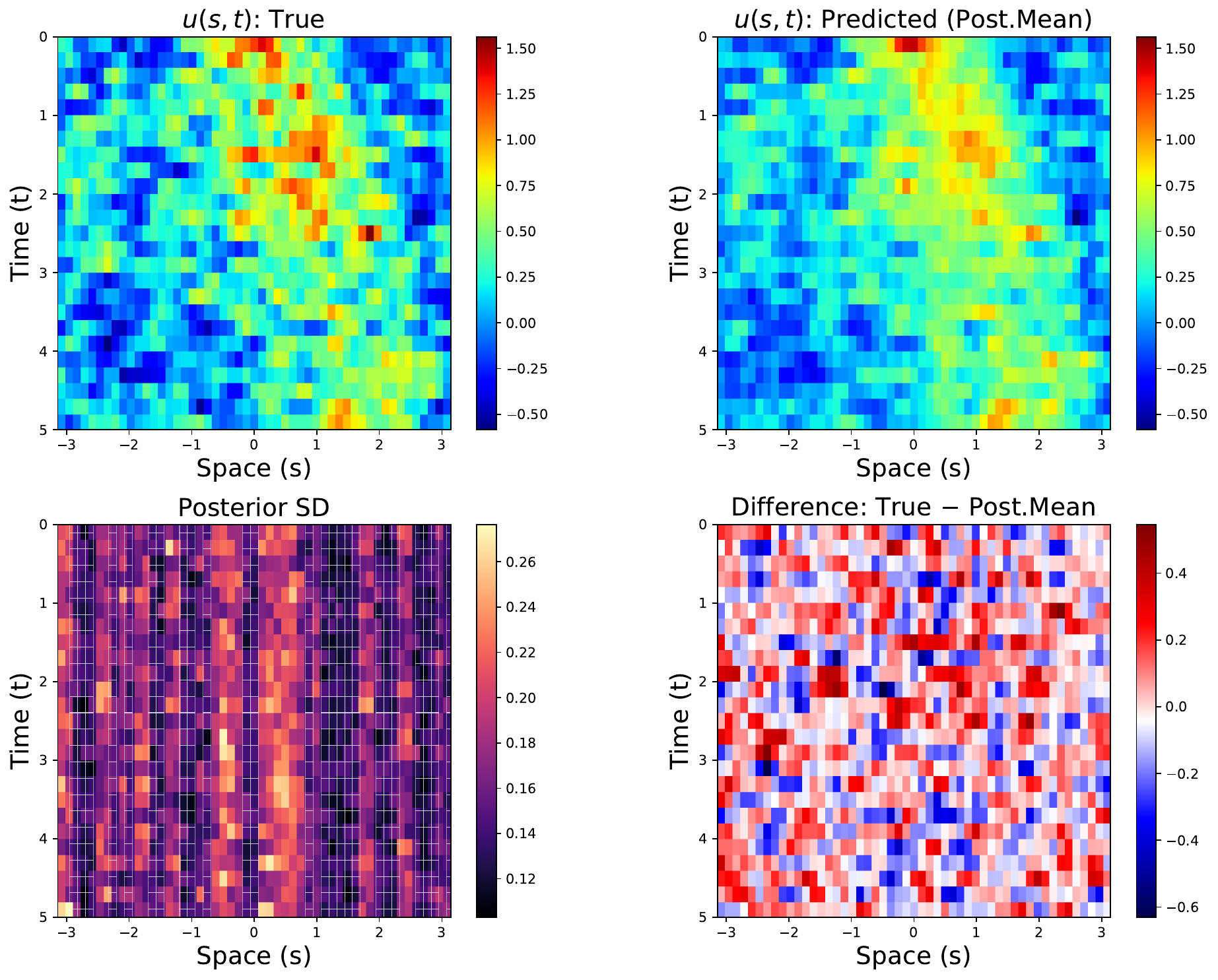}
  \caption{Top Left: True simulated $\bm{u}_t$; Top Right: Predicted posterior mean $\bm{u}_t$; Bottom Left: Posterior standard deviation, with white rectangles indicating locations where data are observed; Bottom Right: Difference between the true $\bm{u}_t$ and the posterior mean.}
    \label{fig6_supple}
\end{figure}
\noindent where we set $\mu_d=\mu_{\delta}=\mu_{\ell}=0,\gamma_d=\gamma_{\delta}=\sigma^2_{\ell}=1$, and truncation bounds $\ell_{d}=0.1,u_d=0.3, \ell_{\delta}=0.02, u_{\delta}=0.1, \ell_{\ell}=0.05,$ and $u_{\ell}=0.2$, in order to assign relatively informative priors for $\sigma^2_d$, $\sigma^2_{\delta}$, and $\ell_{\delta}$. For $\bm{\beta}$, $\lambda$, and neural network weight and bias parameters $\bm{\theta}_W$, we set $\bm{\mu}_{\bm{\beta}}=\bm{\mu}_W=\bm{0}$, $c_W=\sigma^2_{\lambda}=1$, $\mu_{\lambda}=0$, and $c_{\bm{\beta}}=10$. The neural network for ${\bm{u}}_{NN,t}$ consists of $L=3$ hidden layers, each with 16 hidden units. For PDE residuals, boundary, and initial condition data, we consider
\begin{align}
[D_r|\sigma^2_r,\bm{\theta}_W,\lambda]&=\text{N}(0;[\frac{\partial v_{NN}} {\partial t}+ v_{NN}\frac{\partial v_{NN}}{\partial s} - \lambda \frac{\partial ^2 v_{NN}}{{\partial s^2}}](s, t; \bm{\theta}_W),\sigma^2_r), \quad (\text{Burgers' equation})  \nonumber\\
[D_{bc+}|\sigma^2_{bc},\bm{\theta}_W]&=\text{N}(0;v_{NN}(\pi,t; \bm{\theta}_W),\sigma^2_{bc}),\quad (\text{Right boundary condition})  \nonumber \\
[D_{bc-}|\sigma^2_{bc},\bm{\theta}_W]&=\text{N}(0;v_{NN}(-\pi,t; \bm{\theta}_W),\sigma^2_{bc}),\quad (\text{Left boundary condition}) \nonumber \\
[D_{ic}|\sigma^2_{ic},\bm{\theta}_W]&=\text{N}(\exp(-s^2); v_{NN}(s,0; \bm{\theta}_W), \sigma^2_{ic}), \quad (\text{Initial condition}) \nonumber, 
\end{align}
where $\sigma^2_r = 0.05^2$ and $\sigma^2_{bc} = \sigma^2_{ic} = 0.01^2$. 

After specifying the prior distributions and the neural network architecture, we draw 2,000 posterior samples using HMC with the No-U-Turn Sampler \citep{10.5555/2627435.2638586}, discarding the first 1,000 as burn-in and using a tree depth of 10. This is implemented using the ``NumPyro" \citep{phan2019composable} and ``JAX" \citep{jax2018github} packages in Python and takes approximately 45 minutes on an Apple MacBook M4 Pro with 128GB RAM.

Figure \ref{fig3_supple}  presents violin plots for the parameters, excluding the neural network weights and biases $\bm{\theta}_W$, for which the credible intervals contain the true values.
Figure \ref{fig4_supple} show comparisons between $v(s,t)$ and the posterior mean of $v_{NN}(s,t)$, showing that the posterior mean closely aligns with the true $v(s,t)$ despite a substantial amount of missing data. Figure \ref{fig5_supple} compares the true $\bm{\delta}_t$ with its predicted values, indicating that predicting the discrepancy term is more challenging subject to more uncertainty. Nevertheless, the posterior mean (dashed line in the spatial and temporal slice plots) provides a reasonably good match to the true process (solid line).

Figure \ref{fig6_supple} shows the posterior mean of the overall process $\bm{u}_t$. While there are minor deviations from the true $\bm{u}_t$, the posterior mean successfully captures the underlying spatio-temporal pattern, with the posterior standard deviation being higher in locations where data are missing. This demonstrates the flexibility of the model.

\bibliographystyle{apalike-ejor}
\bibliography{bibliography}

\end{document}